\documentclass[aps,prd,reprint,superscriptaddress,nofootinbib]{revtex4-2}
\usepackage[x11names]{xcolor}
\usepackage{lipsum}
\usepackage{scalerel}
\usepackage{amssymb}
\usepackage{mathtools}
\usepackage{amsmath}
\usepackage{braket}
\usepackage{booktabs}
\usepackage{multirow}
\usepackage{makecell}
\usepackage{hyperref}
\hypersetup{
  colorlinks,
  linkcolor={Brown4},
  citecolor={PaleGreen4},
  urlcolor={DodgerBlue2}
}

\def\app#1#2{%
  \mathrel{%
    \setbox0=\hbox{$#1\sim$}%
    \setbox2=\hbox{%
      \rlap{\hbox{$#1\propto$}}%
      \lower1.1\ht0\box0%
    }%
    \raise0.25\ht2\box2%
  }%
}
\def\approxprop{\mathpalette\app\relax}

\begin{document}

\title{Testing screened scalar-tensor theories of gravity with atomic clocks}

\author{Hugo Lévy}
\affiliation{DPHY, ONERA, Universit\'{e} Paris Saclay
F-92322 Ch\^{a}tillon - France}
\affiliation{Sorbonne Universit\'{e}, CNRS, UMR 7095, Institut d’Astrophysique de Paris, 98 bis bd Arago, 75014 Paris, France}
\author{Jean-Philippe Uzan}
\affiliation{Sorbonne Universit\'{e}, CNRS, UMR 7095, Institut d’Astrophysique de Paris, 98 bis bd Arago, 75014 Paris, France}
\affiliation{Center for Gravitational Physics and Quantum Information, Yukawa Institute for Theoretical Physics, Kyoto University, 606-8502, Kyoto, Japan}

\begin{abstract}
In any scalar-tensor theory of gravity exhibiting a screening mechanism, the fifth force mediated by the scalar field is dynamically suppressed at sub-Solar system scales, allowing it to pass existing tests of gravity. As a result, a major research effort has been carried out over the past decades to `outsmart' screened scalars in this game of hide-and-seek. While most of such tests rely on fifth force effects, one should keep in mind that the latter are by no means the only physical feature of scalar-tensor gravity. In particular, this article investigates the possibility of testing screened scalar-tensor models by means of gravitational redshift measurements performed with atomic clocks. Upon deriving the expression for the redshift in this framework, we propose a thought experiment for testing the chameleon model by clock comparisons, which guides us towards more realistic experimental setups, in the laboratory and in space. We find that currently unconstrained regions of the chameleon parameter space could be ruled out by future redshift experiments.
\end{abstract}

\maketitle


\section{Introduction}
\label{sec:introduction}

Scalar-tensor theories of gravity are undeniably one of the most natural and resilient extensions of General Relativity (GR). In this class of models, gravity is mediated by both a rank-2 tensor field \textemdash \ the metric \textemdash \ and a scalar field. The addition of this extra scalar degree of freedom in the gravitational sector, with respect to GR, allows for a wide range of different phenomenologies. The scalar field can indeed be made to play various roles depending on the underlying physical motivations: from driving the late-time universe's accelerated expansion~\cite{copeland-2006, joyce-2015} to dark matter candidates~\cite{peebles-2000, uldm-2017, symmetron-dm-2019}. Scalar fields are also invoked in some attempts to alleviate the Hubble tension~\cite{pitrou-uzan-2024, hogas-2023}. Finally, they are a generic prediction of fundamental theories involving extra dimensions, e.g. Kaluza\textendash Klein theories or string theories in the low energy limit~\cite{uzan-2024, kaluza-klein-gravity-1997, damour-polyakov-1994, gasperini-2001}.

One of the most conspicuous features of scalar-tensor theories of gravity are \textit{fifth forces}, which arise most notably when the scalar field is conformally and universally coupled to the matter sector while remaining minimally coupled to the metric. In theories with screening mechanisms, fifth forces can be greatly mitigated in order to pass existing tests of gravity without compromising the very \textit{raisons d'être} of such theories (e.g. explain cosmic acceleration). In effect, screening mechanisms are nothing but well-chosen nonlinearities in the field equations that dynamically suppress fifth forces at sub-Solar system scales. They include the symmetron~\cite{symmetron-2010, symmetron-2011}, the Damour\textendash Polyakov~\cite{damour-1994-least-coupling, damour-polyakov-2011}, the chameleon~\cite{chameleon-KhouryWeltmanPRD, chameleon-brax-2004} or the Vainshtein~\cite{vainshtein-1, vainshtein-2, vainshtein-3} mechanisms.

Despite being advertised as convenient ways to hide fifth forces in the laboratory and in the Solar system, scalar-tensor models featuring a screening mechanism do nonetheless predict deviations from GR in these environments, no matter how small. In this regard, a major research effort has been carried out over the past decades to find new ways of constraining screened scalar-tensor models. Interestingly, the vast majority of proposed and conducted experiments seem to rely almost exclusively on fifth force effects~\cite{burrage-review-2018, mpb-2021, uzan-2020}. Direct fifth force searches include torsion balance experiments~\cite{adelberger-2003, kapner-2007, upadhye-2012, upadhye-2013}, Casimir force tests~\cite{casimir-tests-2015, casimir-tests-2023} or levitated force sensors~\cite{yin-2022}. Several other tests rely on the fifth force experienced by some test bodies, albeit in a more subtle way. This is notably the case of experiments based on atom interferometry, which made a genuine breakthrough in terms of constraints on the viable parameters of the chameleon or symmetron models~\cite{burrage-2015, hamilton-2015, jaffe-2017, sabulsky-2019}. In these experiments, atoms are put in a superposition of states that travel along different paths. The phase difference accumulated by the wavefunction between the two paths can then be detected as an interference pattern when they are merged. This phase difference being proportional to the total acceleration undergone by the atoms~\cite{peters-2001, hamilton-2015}, such experiments make it possible to constrain scalar fifth forces. On another note, most astrophysical tests also leverage the modified dynamics of massive bodies. For instance, fifth force effects directly relate to the peculiar velocities of galaxies, which can in turn leave a measurable imprint in redshift space~\cite{xu-2015, bose-2016}.

Nonetheless, fifth forces are not the only \textit{measurable} effect inherent to scalar-tensor gravity. Scalar radiation, e.g. in pulsating stars~\cite{silvestri-2011, upadhye-monopole-radiation-2013, dima-2021} or in compact binary systems~\cite{zhang-radiation-compact-binary-2017, zhang-radiation-compact-binary-2022}, is yet another physical manifestation of the scalar field. Back to the laboratory, experiments based on the measurement of atomic energy levels (hydrogen and muonium)~\cite{brax-burrage-2011, brax-davis-2023} or cold bouncing neutrons~\cite{brax-pignol-2011, brax-pignol-2013, jenke-2014, cronenberg-2018} also differ from fifth force searches. In these scenarios, the scalar field behaves as a potential that perturbs the Hamiltonian $\mathcal{H}$ of the system at stake as~\cite{burrage-review-2018, brax-davis-2023}
\begin{equation}
\delta \mathcal{H} \sim m \ln \bigl[ \Omega (\phi) \bigr] \, ,
\label{eqn:perturbed-hamiltonian}
\end{equation}
$m$ being either the mass of the electron or that of the neutron and $\Omega$ the conformal factor that relates the Jordan-frame metric to the Einstein-frame metric [see Eq.~(\ref{eqn:weyl-transform})], which is a function of the scalar field $\phi$. By the same token, experiments based on neutron interferometry depend on the integral $\int  \ln \Omega$ along the neutrons' paths~\cite{brax-neutron-interferometry-2014, lemmel-2015, fischer-2024}. For the sake of clarity, we say that deviations from GR of the form~(\ref{eqn:perturbed-hamiltonian}) are caused by \textit{potential effects}.

Remarkably, a distinction can be drawn between tests based on potential effects \textit{vs} fifth force searches. On the one hand, tests based on potential effects \textemdash \ e.g. measuring the difference between two energy levels of hydrogen-like systems \textemdash \ depend on $\ln [\Omega(\phi) ]$. On the other hand, the relevant quantity in fifth force searches is $\boldsymbol{\nabla} [\ln \Omega (\phi)]$. Consequently, the two kinds of tests are arguably fundamentally different in nature. Moreover, it should be noted that current potential-based tests all have in common the fact that they involve a quantum-mechanical description of the \textit{probe} being used \textemdash \ namely through the quantization of energy levels and the interference of matter waves.\footnote{For the sake of clarity, it is also worth noting that the scalar field effect in those setups is purely classical, as opposed to other tests in which its effects are truly quantum, e.g. one-loop contributions to the electron's magnetic moment \cite{brax-2018}.\label{fn:classical-probe}} This leads to a possible classification of existing tests as depicted in Table~\ref{tab:tests-classification}.

\begin{table}[t]
    \centering
    \begin{ruledtabular}
    \begin{tabular}{ccc}
      & \makecell{$5^{\mathrm{th}}$\,force effects\\``$\boldsymbol{\nabla} [\ln \Omega (\phi)]$''} & \makecell{Potential effects\\``$\ln \Omega(\phi)$''} \\\midrule
      \makecell{Classical \\ probe} & \footnotesize\makecell{torsion balance~\cite{adelberger-2003, kapner-2007, upadhye-2012, upadhye-2013},\\galaxy dynamics~\cite{xu-2015, bose-2016}} & \footnotesize\makecell{gravitational\\redshift} \\[10pt]
      \makecell{Quantum \\ probe} & \footnotesize\makecell{atom \\ interferometry~\cite{burrage-2015, hamilton-2015, jaffe-2017, sabulsky-2019}} & \footnotesize\makecell{atomic levels~\cite{brax-burrage-2011, brax-davis-2023},\\bouncing neutron~\cite{brax-pignol-2011, brax-pignol-2013, jenke-2014, cronenberg-2018}}
    \end{tabular}
    \end{ruledtabular}
    \caption{Classification of tests of screened scalar-tensor models based on: (\textit{i}) whether they hinge on fifth force or potential effects, and (\textit{ii}) whether they require a quantum-mechanical description of the probe at stake or not.\footref{fn:classical-probe} Some actual tests are provided as examples, but the list is by no means exhaustive. The present work extends this landscape by focusing on tests based on the gravitational redshift using clocks, where the latter can \textit{a priori} be treated as classical objects.}
    \label{tab:tests-classification}
\end{table}

This article leaves the quantum realm and examines the phenomenon of \textit{gravitational redshift} (or equivalently, \textit{gravitational time-dilation}) in the framework of scalar-tensor theories. This purely classical effect is common to all metric theories of gravity, and notably manifests itself in the decrease of the frequency of light as it climbs out of a gravitational potential. As for laboratory experiments based on potential effects (see Table~\ref{tab:tests-classification}), the scalar field contributes to the total gravitational redshift through the spacetime dependent quantity $\Omega(\phi)$ [see Eq.~(\ref{eqn:gravitational-redshift-st})]. Unlike these laboratory experiments however, which all involve quantum phenomena, the gravitational redshift is a classical effect\footnote{In this regard, this work mainly focuses on the $00$-component of the Jordan-frame metric tensor rather than on some perturbed Hamiltonian [see Eq.~(\ref{eqn:perturbed-hamiltonian})]. The two interpretations are nonetheless equivalent: the former is best understood in the Jordan frame while the latter relies on computations conducted in the Einstein frame.} and can be measured across large distances using high precision clocks (see e.g. Refs.~\cite{clock-geodesy-2018, galileo1, galileo2, takamoto-2020}). Indeed, clocks can \textit{a priori} be treated as classical objects \textemdash \ although in fine the operating principle of atomic clocks is largely based on quantum mechanics. This opens new venues for testing scalar-tensor models. Indeed, while fifth force searches generally try to maximize ${\boldsymbol{\nabla} [\ln \Omega(\phi)] \propto \boldsymbol{\nabla} \! \phi}$ (see e.g. Ref.~\cite{briddon-2024}) or to disentangle it from Newtonian gravity (see e.g. Refs.~\cite{upadhye-2012, burrage-2016, uzan-2020, mypaper-geodesy-2024}), the constraining power of redshift measurements is intimately related to the variations of $\Omega(\phi)$ from one spatial location to another. As we show in this paper, this leads to radically different experimental designs.

Of course, the fact that the gravitational redshift is altered in scalar-tensor theories with respect to GR's predictions is well-known, and the idea of leveraging this effect to constrain them in turn, is not new. Ref.~\cite{hughes-1990} establishes a rigorous derivation of the anomalous redshift arising from vector and scalar fields non-minimally coupled to matter in the Einstein frame. The seminal article on the chameleon model~\cite{chameleon-KhouryWeltmanPRD} even mentions in Sec.~\textsc{viii} the possibility of deriving constraints from the Vessot\textendash Levine bound~\cite{vessot-levine-1980}, where the authors argue that chameleons comfortably satisfy this bound. Ref.~\cite{minazzoli-2013} also underlines the fact that scalar-tensor theories predict a measured redshift different from that given in GR.

Given all the above, the goal of the present article is to assess the possibility of testing \textit{screened} scalar-tensor models by means of redshift-based experiments involving clocks. In this perspective, we provide a pedagogical derivation of the redshift, both in the Jordan and Einstein frames, with the aim of clarifying any possible source of confusion. Following on from our previous work~\cite{mypaper-geodesy-2024}, we specifically focus on the chameleon model with positive exponent and endeavor to show that precise redshift measurements could reveal the presence of such a scalar field. We imagine an experiment in which atomic clocks are placed in boxes filled with materials of different densities, the goal being to immerse the clocks in different scalar field values. The greater the scalar field contrast, the greater its contribution to the total redshift and the tighter the derived constraints on the model parameters. Such a \textit{gedankenexperiment} has some obvious limitations, which we try to circumvent in more realistic experimental designs. In this endeavor, we rely either  on existing analytical approximations of the chameleon field profile or solve the field's equation numerically with \textit{femtoscope}\footnote{\url{https://github.com/onera/femtoscope}}~\cite{mypaper-nonlinear-kg-fem-2022}. The specific scripts used for this study are publicly accessible~\cite{companion-code}. Singularly, the experimental designs we present would not be relevant if it were not for the screening mechanism. It is precisely the nonlinear nature of the field's equation that allows it to vary significantly between two neighboring space locations, which is desirable in our case. This leads us to propose novel ideas for experimental tests, both in the laboratory and in space.

This article is organized as follows. In Sec.~\ref{sec:stt-definitions-notations}, we  recall the theoretical foundations for the class of conformally coupled scalar-tensor models. In this framework, we give in Sec.~\ref{sec:gravitational-redshift-stt} the redshift expression and study its Newtonian limit in the special case of the chameleon model. Building on these insights, we propose a thought experiment in Sec.~\ref{sec:thought-experiment} for testing the chameleon model with atomic clocks. In particular, we lay emphasis on the fact that, despite satisfying Local Position Invariance (LPI), chameleon gravity can nonetheless be distinguished from GR in redshift measurements. The orders of magnitude derived there being competitive with current bounds on the model, we go a step further in Sec.~\ref{sec:more-realistic-experiments} in order to assess whether they hold in more realistic setups \textemdash \ in the laboratory on Earth, and in space with satellites. Finally, we conclude in Sec.~\ref{sec:conclusion}.

\section{Scalar-tensor theories: definitions and notations}
\label{sec:stt-definitions-notations}

In this article, we use the signature convention ${(-, \, + , \, + , \, +)}$ and work in natural units, for which ${c = \hbar = 1}$. Greek indices ${(\mu, \, \nu , \, \rho , \, \sigma , \, \text{etc.})}$ run from 0 to 3 while Latin indices ${(i , \, j , \, k , \, \text{etc.})}$ run from 1 to 3. We shall focus on scalar-tensor models whose action can be put in the form
\begin{equation}
S = S_{\mathrm{EH}} + S_{\phi} + S_{\mathrm{mat}} \, ,
\label{eqn:action-ef}
\end{equation}
where
\begin{subequations}
\begin{align}
S_{\mathrm{EH}} &= \frac{M_{\mathrm{Pl}}^2}{2} \int \mathrm{d}^4 x \sqrt{-g} R \, , \label{subeqn:einstein-hilbert-action} \\[3pt]
S_{\phi} &= - \int \mathrm{d}^4 x \sqrt{-g} \biggl[ \frac{1}{2} g^{\mu \nu} \partial_{\mu} \phi \, \partial_{\nu} \phi + V(\phi) \biggr] \, , \label{subeqn:scalar-field-action} \\[3pt]
S_{\mathrm{mat}} &= \int \mathrm{d}^4 x \sqrt{-\tilde{g}} \, \mathcal{L}_{\mathrm{mat}} \bigl( \tilde{g}_{\mu \nu} , \, \psi_{\mathrm{mat}}^{\scaleto{(i)}{5.5pt}} \bigr) \, . \label{subeqn:matter-action}
\end{align}
\label{eqn:action-parts}%
\end{subequations}
Eq.~(\ref{subeqn:einstein-hilbert-action}) is the standard Einstein\textendash Hilbert action, featuring the reduced Planck mass $M_{\mathrm{Pl}} = 1 / \sqrt{8 \pi G}$, the Ricci scalar $R$ constructed from the Einstein-frame metric $g_{\mu \nu}$. Eq.~(\ref{subeqn:scalar-field-action}) is the action of the scalar field $\phi$,\footnote{With these conventions, $\phi$ can be expressed in electronvolts.} with a canonical kinetic term and potential $V$. Eq.~(\ref{subeqn:matter-action}) denotes the matter action, where all matter fields $\psi_{\mathrm{mat}}^{\scaleto{(i)}{5.5pt}}$ are universally coupled to the Jordan-frame metric $\tilde{g}_{\mu \nu}$. The latter is chosen to be related to the Einstein-frame metric through a Weyl transformation, reading
\begin{equation}
\tilde{g}_{\mu \nu} = \Omega^2 (\phi) g_{\mu \nu} \, ,
\label{eqn:weyl-transform}
\end{equation}
for some real conformal factor function $\Omega$. Other Jordan-frame quantities are also denoted with a tilde throughout this article, except for the Jordan-frame counterpart of $\phi$ which is denoted $\varphi$ (see Appendix~\ref{subsec:field-equations-jf} for details).

In the following, we compile the fields and geodesic equations \textemdash \ together with their Newtonian limit \textemdash \ in the Einstein frame, as they will prove useful in the subsequent sections. For the sake of completeness, additional derivations, including Jordan-frame expressions, can be found in Appendix~\ref{app:additional-derivations}.

\subsection{Fields equations in the Einstein frame}
\label{subsec:field-equations-newtonian-limits}

The fields equations are obtained by varying the action given by Eqs.~(\ref{eqn:action-ef}\textendash\ref{eqn:action-parts}) with respect to $g^{\mu \nu}$ and $\phi$, yielding
\begin{align}
G_{\mu \nu} \equiv R_{\mu \nu} - \frac{1}{2} R g_{\mu \nu} &= \frac{1}{M_{\mathrm{Pl}}^2} \bigl( T_{\mu \nu} + T_{\mu \nu}^{\scaleto{(\phi)}{5pt}} \bigr) \label{eqn:metric-eqn-ef} \\[3pt]
\Box \phi \equiv g^{\mu \nu} \nabla_{\!\mu} \! \nabla_{\! \nu} \phi &= \frac{\mathrm{d} V}{\mathrm{d} \phi} - \frac{\mathrm{d} \ln \Omega}{\mathrm{d} \phi} T  \label{eqn:scalar-field-eqn-ef}
\end{align}
In the above, we have introduced the stress-energy tensor of matter
\begin{equation}
T_{\mu \nu} \equiv \frac{-2}{\sqrt{-g}} \frac{\delta S_{\mathrm{mat}}}{\delta g^{\mu \nu}}
\label{eqn:stress-energy-tensor-ef}
\end{equation}
with trace $T = g^{\mu \nu} T_{\mu \nu}$, as well as the scalar field stress-energy tensor
\begin{equation}
\begin{split}
T_{\mu \nu}^{\scaleto{(\phi)}{5pt}} &\equiv \frac{-2}{\sqrt{-g}} \frac{\delta S_{\phi}}{\delta g^{\mu \nu}} \\
&= \partial_{\mu} \phi \, \partial_{\nu} \phi - \frac{1}{2} g_{\mu \nu} g^{\rho \sigma} \partial_{\rho} \phi \, \partial_{\sigma} \phi - g_{\mu \nu} V(\phi) \, .
\end{split}
\label{eqn:stress-energy-tensor-scalar-ef}
\end{equation}

To get the Newtonian limit of Eqs.~(\ref{eqn:metric-eqn-ef}\textendash\ref{eqn:scalar-field-eqn-ef}), we assume that the Einstein-frame metric can be expanded about the Minkowski metric $\eta_{\mu \nu}$ as ${g_{\mu \nu}  = \eta_{\mu \nu} + h_{\mu \nu}}$, with $|h_{\mu \nu}| \ll 1$. Choosing a coordinate system $(t , \, x^i)$ so that ${\eta_{\mu \nu} = \mathrm{diag}(-1 , \, 1 , \, 1 , \, 1)}$, gauge freedom allows us to put the metric in the form
\begin{equation}
ds^2 \equiv g_{\mu \nu} \mathrm{d}x^{\mu} \mathrm{d}x^{\nu} \! = -(1+ 2 \Phi) \mathrm{d}t^2 + g_{ij} \mathrm{d}x^i \mathrm{d}x^j ,
\label{eqn:newton-gauge-ef}
\end{equation}
featuring a potential $\Phi$, with $|\Phi| \ll 1$. For the Newtonian limits in both frames to be consistent with one another, we further have to assume that the conformal factor $\Omega$ is close to unity, that is\footnote{One may find it helpful to keep in mind the common form for the conformal factor $\Omega(\phi) = \exp(\alpha \phi)$, where $\alpha$ is just a coupling constant. In that case $\omega(\phi) \sim \alpha \phi$.}
\begin{equation}
\Omega(\phi) = 1 + \omega(\phi) \, , \qquad \text{with} \quad |\omega(\phi)| \ll 1 \, .
\label{eqn:conformal-factor-approximation}
\end{equation}
The need for this additional hypothesis is made clearer in Appendix~\ref{subsec:jf-newtonian-limit}. The 00-component of Eq.~(\ref{eqn:metric-eqn-ef}), for static configurations, then becomes
\begin{equation}
2 M_{\mathrm{Pl}}^2 \Delta \Phi = \rho - 2 V(\phi) \, ,
\label{eqn:metric-eqn-newtonian-limit-ef}
\end{equation}
while the scalar field equation~(\ref{eqn:scalar-field-eqn-ef}) boils down to
\begin{equation}
\Delta \phi = \frac{\mathrm{d} V}{\mathrm{d} \phi} + \frac{\mathrm{d} \ln \Omega}{\mathrm{d} \phi} \rho \, .
\label{eqn:scalar-field-eqn-newtonian-limit-ef}
\end{equation}
Here, it should be noted that condition~(\ref{eqn:conformal-factor-approximation}) allows us to drop powers of the conformal factor multiplying already-small quantities. In that respect, ${\omega(\phi) \rho}$ constitutes a higher-order term, so that one can approximate $\tilde{\rho} \simeq \rho$ (see Appendix~\ref{subsec:stress-energy-tensors-both-frames}). The Laplacian is formally defined as $\Delta \equiv g^{ij} \partial_i \partial_j$ but will be approximated as $\Delta = \delta^{ij} \partial_i \partial_j$.

\subsection{Geodesic equation}
\label{subsec:geodesic-equation}

It is not uncommon to read the Jordan frame being referred to as the `physical frame' in the literature \cite{cosmological-equivalence-frames, gef-polarski-2001, gef-2011}. It is the frame in which the matter action takes its standard form: the stress-energy tensor is covariantly conserved and all particle physics' properties (e.g. masses, cross sections, decay rates, etc.) can be computed `as usual', without having to care about the spacetime dependence of the scalar field $\varphi$. In particular, matter test particles in free fall move along geodesics of the Jordan-frame metric. Denoting ${\tilde{u}^{\mu} = \mathrm{d}x^{\mu} / \mathrm{d} \tilde{\tau}}$ the 4-velocity of such a test particle \textemdash \ where ${\mathrm{d \tilde{\tau}} \equiv - d\tilde{s}}$ [see Eq.~(\ref{eqn:newton-gauge-jf})] \textemdash, we have
\begin{equation}
\tilde{u}^{\alpha} \tilde{\nabla}_{\! \alpha} \tilde{u}^{\mu} = 0 \, , \qquad \text{with} \quad \tilde{g}_{\alpha \beta} \tilde{u}^{\alpha} \tilde{u}^{\beta} = -1 \, .
\label{eqn:geodesic-eqn-jf}
\end{equation}
In the Newtonian limit, Eq.~(\ref{eqn:geodesic-eqn-jf}) takes the familiar form
\begin{equation}
\frac{\mathrm{d}^2 x^i}{\mathrm{d}t^2} = - \partial_i \tilde{\Phi} \, ,
\label{eqn:geodesic-newtonian-limit-jf}
\end{equation}
where the Jordan-frame potential $\tilde{\Phi}$ is defined by Eq.~(\ref{eqn:newton-gauge-jf}) in Appendix~\ref{subsec:jf-newtonian-limit}.

It is possible to rewrite the geodesic equation~(\ref{eqn:geodesic-eqn-jf}) in a form involving only Einstein-frame quantities. One finds that geodesics of the Jordan-frame metric do not coincide with those of the Einstein-frame metric, instead
\begin{equation}
u^{\alpha} \nabla_{\! \alpha} u^{\mu} = - \perp^{\mu \nu} \frac{\mathrm{d} \ln \Omega}{\mathrm{d} \phi} \partial_{\nu} \phi \, ,
\label{eqn:geodesic-eqn-ef}
\end{equation}
with $\perp^{\mu \nu} = g^{\mu \nu} + u^{\mu} u^{\nu}$. Therefore, from the Einstein-frame perspective, everything happens as if the particle were subjected to a fifth force. In the Newtonian limit, Eq.~(\ref{eqn:geodesic-eqn-ef}) reduces to
\begin{equation}
\frac{\mathrm{d}^2 x^i}{\mathrm{d}t^2} = - \partial_i \Phi - \frac{\mathrm{d}\ln \Omega}{\mathrm{d}\phi} \partial_i \phi = - \partial_i \bigl( \Phi + \ln \Omega \bigr) \, .
\end{equation}

The case of null geodesics is discussed in Appendix~\ref{subsec:null-geodesics}. In particular, conformal transformations leave null geodesics invariant.

\section{Gravitational redshift in scalar-tensor theories}
\label{sec:gravitational-redshift-stt}

This section provides, under very general assumptions, the expression of the redshift in scalar-tensor theories. Emphasis is laid on \textit{measurable quantities} \textemdash \ in that respect, the expression~(\ref{eqn:gravitational-redshift-st}) thereby obtained is put into perspective with (\textit{i}) the cosmological setting, and (\textit{ii}) the usual form of parameterized redshift violations. Finally, we show that the chameleon model could be quite sensitive to redshift tests in some specific cases, precisely because of their inherent nonlinear character. In this endeavor, we frequently refer to the equations derived back in Sec.~\ref{sec:stt-definitions-notations}.

\subsection{Derivation of the measured redshift in scalar-tensor theories}
\label{subsec:redshift-derivation-stt}

Suppose one observer, the \textit{emitter}, sends a photon to another observer, the \textit{receiver}. We make the two following assumptions:
\begin{enumerate}
    \item The spatial coordinates of the two observers, $(x^i)_{\mathrm{em}}$ and $(x^i)_{\mathrm{rec}}$, remain constant throughout the experiment.
    \item The metric $\tilde{g}_{\mu \nu}$ is stationary, meaning that it does not depend upon the $x^0$ coordinate, i.e. $\partial_0 \tilde{g}_{\mu \nu} = 0$. Consequently, the scalar field $\phi$ will also be assumed stationary [which has already been assumed in Eq.~(\ref{eqn:scalar-field-eqn-newtonian-limit-ef})].
\end{enumerate}
Then, we show in Appendix~\ref{app:redshift-expression} that the gravitational redshift $z$ of the photon in the framework of scalar-tensor theories is given by
\begin{equation}
	1+z = \sqrt{\frac{(\tilde{g}_{00})_{\mathrm{rec}}}{(\tilde{g}_{00})_{\mathrm{em}}}} = \frac{\Omega_{\mathrm{rec}}}{\Omega_{\mathrm{em}}} \sqrt{\frac{(g_{00})_{\mathrm{rec}}}{(g_{00})_{\mathrm{em}}}} \, .
\label{eqn:gravitational-redshift-st}
\end{equation}

This formula sheds light on the dependence of the redshift on the scalar field $\phi$. On the one hand, the Einstein-frame metric coefficient $g_{00}$ intricately depends on $\phi$ through Eq.~(\ref{eqn:metric-eqn-ef}). Taking the Newtonian limit of this equation helps clarify this dependence \textemdash \ Eq.~(\ref{eqn:metric-eqn-newtonian-limit-ef}) indeed shows that the potential $\Phi = -(g_{00}+1)/2$ obeys a Poisson equation where the scalar potential $V(\phi)$ is part of the source term alongside $\rho$. On the other hand, the presence of the conformal factor $\Omega$ is somewhat easier to interpret as it is merely a function of the scalar field. Therefore, $z$ is a \textit{measurable} quantity that (a priori) depends on the scalar field's amplitude at the emission and reception spacetime events. In Sec.~\ref{subsec:redshift-focus-chameleon}, we shall expand Eq.~(\ref{eqn:gravitational-redshift-st}) in the framework of the chameleon model and study the corresponding Newtonian limit.

\subsection{Side note: cosmological setting}
\label{subsec:derivation-cosmological-setting}

The above derivation of the redshift formula~(\ref{eqn:gravitational-redshift-st}) in scalar-tensor theories also applies to the cosmological setting. In the Einstein frame, the spatially flat \textsc{flrw} line element reads
\begin{equation}
ds^2 = - \mathrm{d}t^2 + a^2(t) \delta^{ij} \mathrm{d}x^i \mathrm{d}x^j \, ,
\label{eqn:flrw-metric-ef}
\end{equation}
featuring the scale factor $a(t)$. In particular, we retrieve the expression
\begin{equation}
1 + z = \frac{\Omega_{\mathrm{rec}}}{\Omega_{\mathrm{em}}} \frac{a_{\mathrm{rec}}}{a_{\mathrm{em}}}
\end{equation}
giving the redshift of a distant object in the sky (in scalar-tensor gravity).\footnote{One may rightly object that the \textsc{flrw} metric is not stationary and so the redshift derivation we just conducted does not apply. Nevertheless, this stationarity assumption is stronger than needed. It is in fact sufficient to note that $\partial_{\eta}$ is a \textit{conformal Killing vector} on the \textsc{flrw} spacetime manifold, that is tangent to the world lines of the source and of the observer (comoving with the Hubble flow), so that Eq.~(\ref{eqn:killing-vector-property-redshift}) still applies. Here, $\eta$ denotes the conformal time which is related to the coordinate time through $\mathrm{d}t = a(t) \mathrm{d} \eta$ and $a$ is the scale factor. See e.g. Ref.~\cite{redshift-killing-vectors} for more insights into these mathematical considerations.} Therefore, the two following statements hold simultaneously:
\begin{enumerate}
	\item Null geodesics are invariant under conformal transformations in a four-dimensional spacetime. Thus, light-like geodesics of the Jordan-frame metric $\tilde{g}_{\mu \nu}$ coincide with those of the Einstein-frame metric $g_{\mu \nu}$ and massless particles, such as photons, do not `feel' any force from the scalar field (see Appendix~\ref{subsec:null-geodesics}).
	\item The amount by which light emitted from distant objects gets redshifted (through the expansion of the universe) when it eventually gets to us depends explicitly on the scalar field and its cosmological evolution.
\end{enumerate}
This is a common source of confusion, see e.g. Ref.~\cite{veilled-gr-2011}.

\subsection{Gravitational redshift and gravitational potential}
\label{subsec:link-observable-quantities}

\subsubsection{Parameterized redshift tests}
\label{subsubsec:parameterized-redshift-tests}

Experiments that measure the gravitational frequency shift of light usually introduce a dimensionless parameter $\alpha$ to quantify deviations from what is predicted by GR. As such, $\alpha$ is \textit{defined} as
\begin{equation}
    z_{12} = (1+\alpha) \Delta_{12} U ,
\label{eqn:redshit-parametrized-violation}
\end{equation}
where, for two locations, $\Delta_{12}U = U_2 - U_1$. Testing the consistency of $\alpha$ with $0$ is a test of LPI \textemdash \ which is of course embedded in GR. Current upper bounds on $| \alpha |$ are around $10^{-5}$ (see Sec.~\ref{subsec:atomic-clocks} thereafter). In Eq.~(\ref{eqn:redshit-parametrized-violation}), $U$ is either referred to as ``the Newtonian potential'' \cite{chameleon-KhouryWeltmanPRD, aces-proposal-2019}, or as the ``gravitational potential'' \cite{takamoto-2020, manly-2001, radioastron-2023, galileo1, galileo2}. The bothering issue with this designation is that it is unclear how one should actually define and measure it. Along the lines of Will's book~\cite{will-2018}, we define $U$ as the ``gravitational potential whose gradient is related to the test-body acceleration'', i.e. in the Newtonian limit,
\begin{equation}
    \mathbf{a} = -\boldsymbol{\nabla} U \, .
\label{eqn:potential-gradient-definition-acceleration}
\end{equation}
Acceleration $\boldsymbol{\nabla}U$ and redshift $z$ can be measured with accelerometers and clocks respectively, and
\begin{equation}
    \Delta_{12} U = \int_{\mathcal{C}} \mathbf{a} \cdot \mathrm{d}\mathbf{l} \, , \ \forall \, \mathcal{C} \text{ joining points 1 and 2} \, .
\label{eqn:delta-potential-definition}
\end{equation}
Hence, Eqs~(\ref{eqn:redshit-parametrized-violation}\textendash\ref{eqn:delta-potential-definition}) are a check of the consistency between clock comparisons and acceleration measurements. If the separation between point 1 and point 2 is relatively small compared to the characteristic length scale of $U$-variations, Eq.~(\ref{eqn:delta-potential-definition}) can be simplified to $\Delta_{12} U = \mathbf{g} \cdot \mathbf{r}_{12}$, where $\mathbf{g}$ is the gravitational field and $\mathbf{r}_{12}$ is the vector joining the two positions.\footnote{This approximation is performed for laboratory experiments on Earth, see e.g. Refs.~\cite{manly-2001, liu-2024}.} If LPI holds, then $\alpha = 0$. In particular $\alpha_{\scaleto{\mathrm{GR}}{4pt}} = 0$ (at the first post-Newtonian order).

\subsubsection{Newtonian limit in the Jordan frame}
\label{subsubsec:redshift-newtonian-limit-jf}

Let us now show that scalar-tensor theories that fall into the class of models introduced in Sec.~\ref{sec:stt-definitions-notations} also verify $\alpha_{\scaleto{\mathrm{ST}}{4pt}} = 0$. In the Newtonian limit, the Jordan-frame metric can be put in the form~(\ref{eqn:newton-gauge-jf}) with $|\tilde{\Phi}| \ll 1$. Substituting this definition in Eq.~(\ref{eqn:gravitational-redshift-st}) yields, at first order,
\begin{equation}
    z = \sqrt{\frac{1+2\tilde{\Phi}_{2}}{1+2\tilde{\Phi}_{1}}} - 1 \simeq \tilde{\Phi}_{2} - \tilde{\Phi}_{1} = \Delta_{12} \tilde{\Phi} \, ,
\label{eqn:gravitational-redshift-newtonian-jf}
\end{equation}
where we have further re-labeled by 1 and 2 the emission and reception events respectively. There only remains to check that $U \equiv \tilde{\Phi}$. This is actually done by looking at the geodesic equation in the Newtonian limit in the case of a stationary metric \textemdash \ Eq.~(\ref{eqn:geodesic-newtonian-limit-jf}):
\begin{equation}
    \frac{\mathrm{d}^2 x^i}{\mathrm{d}t^2} \simeq - \partial_i \tilde{\Phi} \implies U = \tilde{\Phi} \implies \alpha_{\scaleto{\mathrm{ST}}{4pt}} = 0 \, .
\label{eqn:jf-geodesic-equation-newtonian-limit}
\end{equation}

A few remarks are in order. First, the Jordan frame scalar field $\varphi$ does not appear explicitly in Eq.~(\ref{eqn:jf-geodesic-equation-newtonian-limit}). However, recall that the potential $\tilde{\Phi}$ is obtained through the field equations~(\ref{eqn:metric-eqn-jf}\textendash\ref{eqn:scalar-field-eqn-jf}), which of course depend on the scalar field. In other words, $\tilde{\Phi}$ cannot be considered as the ``Newtonian potential'' in the sense that it does not obey the usual Poisson's equation in the static regime. Instead, Eqs.~(\ref{eqn:metric-eqn-jf}\textendash\ref{eqn:scalar-field-eqn-jf}) remain coupled second-order partial differential equations. Second, finding $\alpha_{\scaleto{\mathrm{ST}}{4pt}} = 0$ should not come as a surprise at all. Indeed, conformally coupled scalar-tensor models belong to the wider class of \textit{metric theories}, which all satisfy the Einstein Equivalence Principle (EEP), including LPI. One may however raise the objection that there is no point in trying to use redshift measurements to constrain scalar-tensor gravity, since in particular the latter satisfies LPI and is thus consistent with all bounds on the parameter $\alpha$. In the face of this argument, we stress that it is not because a given theory satisfies LPI that it cannot be distinguished from GR in redshift experiments, as we will show in Sec.~\ref{sec:thought-experiment}.

\subsubsection{Newtonian limit in the Einstein frame}
\label{subsubsec:redshift-newtonian-limit-ef}

Similarly, we derive the Newtonian limit of the redshift expression~(\ref{eqn:gravitational-redshift-st}) in the Einstein frame. Again, we have to assume condition~(\ref{eqn:conformal-factor-approximation}) for the Newtonian approximations in both frames to be consistent with one another. The Einstein-frame metric is put in the form~(\ref{eqn:newton-gauge-ef}) with ${|\Phi| \ll 1}$. These assumptions allow for the identification ${\tilde{\Phi} \simeq \Phi + \omega(\phi)}$. We then immediately obtain
\begin{equation}
	z = \Delta_{12} \bigl[ \Phi + \omega(\phi) \bigr] \, .
\label{eqn:gravitational-redshift-newtonian-ef}
\end{equation}
The Newtonian limit of the geodesic equation in the Einstein frame is
\begin{equation*}
	\frac{\mathrm{d}^2 \mathbf{x}}{\mathrm{d}t^2} = - \boldsymbol{\nabla} \Phi - \boldsymbol{\nabla} \bigl[ \ln \Omega(\phi) \bigr] \simeq - \boldsymbol{\nabla} \bigl[ \Phi + \omega (\phi) \bigr] \, ,
\end{equation*}
and so we recover the fact that $U = \Phi + \omega(\phi)$ and ${\alpha_{\scaleto{ST}{4pt}} = 0}$, as expected. $\Phi$ and $\phi$ are solutions to Eq.~(\ref{eqn:metric-eqn-newtonian-limit-ef}) and Eq.~(\ref{eqn:scalar-field-eqn-newtonian-limit-ef}) respectively.

All subsequent computations can be conducted in the Einstein frame. Essentially, in order to be able to discuss the redshift, we will have to solve the modified Poisson equation~(\ref{eqn:metric-eqn-newtonian-limit-ef}) and the Klein\textendash Gordon equation~(\ref{eqn:scalar-field-eqn-newtonian-limit-ef}). In particular, the latter does not depend on $\Phi$ and should thus be solved first, yielding $\phi$. Only then can we tackle the modified Poisson equation, because the source term $V(\phi)$ is fully determined after completion of the first step.

\subsection{Focus on the chameleon model}
\label{subsec:redshift-focus-chameleon}

We now focus on the chameleon model with exponent ${n > 0}$, given by the functions
\begin{equation}
	\Omega(\phi) = \exp \left( \frac{\beta \phi}{M_{\mathrm{Pl}}} \right) \quad \text{and} \quad V(\phi) = \Lambda^4 \left( \frac{\Lambda}{\phi} \right)^{\! n} .
\label{eqn:functions-chameleon}
\end{equation}
In the study of fifth force effects, the bare potential function $V$ can be defined up to an additive constant since it only plays a role through its derivatives in computations. Things are different here since $V(\phi)$ appears \textit{as is} in Eq.~(\ref{eqn:metric-eqn-newtonian-limit-ef}). As long as ${\phi \ll M_{\mathrm{Pl}} / \beta}$, assumption~(\ref{eqn:conformal-factor-approximation}) holds and we get $\omega(\phi) = \beta \phi / M_{\mathrm{Pl}}$. With the functions~(\ref{eqn:functions-chameleon}) and neglecting $p \ll \rho$, the scalar field obeys a nonlinear Klein\textendash Gordon equation
\begin{equation}
	\Delta \phi = \frac{\mathrm{d} V_{\mathrm{eff}}}{\mathrm{d} \phi} = \frac{\beta}{M_{\mathrm{Pl}}} \rho - n \frac{\Lambda^{n+4}}{\phi^{n+1}} \, ,
\label{eqn:chameleon-static-eqn}
\end{equation}
where $V_{\mathrm{eff}}(\phi) \equiv V(\phi) + \rho \ln \Omega(\phi)$ is the so-called \textit{effective potential}. The field's value that minimizes this effective potential $\phi_{\mathrm{min}}$ together with the effective mass $m_{\phi}$ are given by
\begin{align}
\phi_{\min} (\rho) &= \left( M_{\mathrm{Pl}} \frac{n \Lambda^{n+4}}{\beta \rho} \right)^{\! \frac{1}{n+1}} \, , \label{eqn:chameleon-effective-min} \\
m_{\phi}^2 (\rho) &= n(n+1) \Lambda^{n+4} \left( \frac{\beta \rho}{n M_{\mathrm{Pl}} \Lambda^{n+4}} \right)^{\! \frac{n+2}{n+1}} \, . \label{eqn:chameleon-effective-mass}
\end{align}
The effective Compton wavelength of the field $\lambda_{\phi}$ is related to the effective mass through $\lambda_{\phi} = 1/m_{\phi}$.

For this specific scalar-tensor model, the redshift expression~(\ref{eqn:gravitational-redshift-newtonian-ef}) becomes
\begin{equation}
	z = \Delta_{12} \left[ \Phi + \frac{\beta \phi}{M_{\mathrm{Pl}}} \right] .
\label{eqn:gravitational-redshift-newtonian-ef-chameleon}
\end{equation}
There, it is already interesting to note that, unlike the chameleonic force which is proportional to the gradient of the scalar field $\boldsymbol{\nabla} \! \phi$, part of the chameleon contribution to the total redshift is proportional to $\Delta_{12} \phi = \phi_2 - \phi_1$. This mere observation has important consequences in terms of choice of experimental designs when it comes to constraining such a model. Maximizing $\| \boldsymbol{\nabla} \! \phi \|$ is undeniably not the same thing as maximizing $\Delta_{12} \phi$.\footnote{The mean value theorem nonetheless establishes a link between these two quantities.} The more intricate $\phi$-dependence through $\Phi$ is examined in more details in Sec.~\ref{subsec:gedankenexperiment}.

For the scalar field to leave a measurable imprint on the total redshift~(\ref{eqn:gravitational-redshift-newtonian-ef-chameleon}), the scalar field must be able to vary significantly from one place to another. In that respect, the chameleon field specifically may actually be a very good candidate. Indeed, Eq.~(\ref{eqn:chameleon-effective-min}) highlights the fact that $\phi_{\min} (\rho) \propto 1 / \rho^{1/(n+1)}$. In particular, $\phi_{\min} (\rho) \to + \infty$ as $\rho \to 0$. Deep inside a medium of constant density $\rho$, ${\phi \sim \phi_{\min} (\rho)}$ provided this medium occupies a large enough spatial region. All in all, this means that the chameleon field should grow to very large values in vast-enough, low-density environments.

Finally, with a view to solve Eq.~(\ref{eqn:chameleon-static-eqn}) numerically, we shall work with its dimensionless counterpart
\begin{equation}
	\alpha_{\scaleto{\mathrm{c}}{3pt}} \bar{\Delta} \bar{\phi} = \bar{\rho} - \bar{\phi}^{-(n+1)} \, .
\label{eqn:chameleon-static-eqn-dimensionless}
\end{equation}
Here, we introduced $L_0$ a characteristic length scale and $\rho_0$ a characteristic density of the problem, and set ${\bar{\Delta} = L_0 \Delta}$, ${\bar{\rho} = \rho / \rho_0}$. Likewise, ${\bar{\phi} = \phi / \phi_0}$ with
\begin{equation}
	\phi_0 = \phi_{\min}(\rho_0) = \left( \frac{n M_{\mathrm{Pl}} \Lambda^{n+4}}{\beta \rho_0} \right)^{\! \frac{1}{n+1}} .
\label{eqn:phi0}
\end{equation}
The dynamics of $\bar{\phi}$ is then entirely controlled by the two dimensionless parameters $n$ and
\begin{equation}
	\alpha_{\scaleto{\mathrm{c}}{3pt}} = \left( \frac{\Lambda M_{\mathrm{Pl}}}{\beta \rho_0 L_0^2} \right) \left( \frac{n M_{\mathrm{Pl}} \Lambda^3}{\beta \rho_0} \right)^{\! \frac{1}{n+1}} ,
\label{eqn:alpha-c}
\end{equation}
effectively reducing the dimension of the number of free parameters from three ${(\beta, \, \Lambda , \, n)}$ to two ${(\alpha_{\scaleto{\mathrm{c}}{3pt}} , \, n)}$. This allows for a much more efficient exploration of the chameleon parameter space. In the remainder of this article, we use ${L_0 = 1 \, \mathrm{m}}$ and ${\rho_0 = 1 \, \mathrm{kg / m^3}}$.

\section{Principle of the experiment and first orders of magnitude}
\label{sec:thought-experiment}

So far, we have derived the redshift formula in scalar-tensor theories and showed its dependence on the scalar field. However, we have yet to show how to translate redshift measurements into actual constraints on the scalar-tensor model at stake. Following on from the previous section, this discussion is illustrated with the example of the chameleon field again. After briefly reviewing the current state of the art in atomic clocks, we propose a thought experiment, underlying more realistic experimental designs, for testing chameleon gravity.

\subsection{Short review on atomic clocks and redshift tests}
\label{subsec:atomic-clocks}

Measuring the gravitational redshift effect on Earth is best achieved by atomic clocks. Indeed, these devices represent the pinnacle of precision timekeeping, playing a critical role in fundamental physics experiments \cite{safronova-review-2018} and underlying the definition of the second in the International System of Units~\cite{dimarcq-2024}. They rely on the ultra-stable atomic transitions to measure time with unparalleled accuracy. Among the most advanced types are optical lattice clocks which probe the optical transitions of trapped ions or atoms with laser light. They achieve relative frequency precisions of $10^{-18}$ and below \cite{nicholson-2015, huntemann-2016, bothwell-2019, brewer-2019, noriaki-2021, aeppli-2024}.

Measuring the gravitational redshift using such devices amounts to comparing the relative frequencies of two clocks placed in different gravitational potentials. The relative uncertainty on such measurements is typically of the same order of magnitude as the systematic uncertainty of the clocks used, i.e. at $O(10^{-18})$ for two independent atomic clocks \cite{takamoto-2020, bacon-2021}. Multiplexed optical lattice clocks can exhibit even lower relative uncertainty on redshift measurements \cite{zheng-2022, zheng-2023}, as low as $7.6 \times 10^{-21}$ \cite{bothwell-2022}, but the atomic ensembles remain very localized in space.

These levels of instability and systematic uncertainty open the way to stringent tests of GR, notably by putting upper bounds on parametrized tests of gravitational redshift (see Sec.~\ref{subsec:link-observable-quantities}). In space, comparing the frequency of hydrogen masers onboard Galileo satellites with eccentric orbits have produced the strongest limits on deviations from the expected redshift \textemdash \ at the $10^{-5}$ level on $\alpha$ defined by Eq.~(\ref{eqn:redshit-parametrized-violation}) \cite{galileo1, galileo2}.\footnote{Note that this new bound is approximately one order of magnitude lower than the one obtained by Vessot--Levine experiment~\cite{vessot-levine-1980}.} The ACES mission \cite{aces-proposal-2019}, to be launched in 2025, aims at improving that bound by one order of magnitude. In terms of prospects, there are other proposals for clocks in space, such as the FOCOS proposal~\cite{focos-proposal-2022}, a space mission that would place a state-of-the-art Yb optical atomic clock in an eccentric orbit around the Earth to reach an uncertainty of $10^{-9}$ on $\alpha$. On Earth, 18-digit-precision frequency comparison between transportable Sr-based optical lattice clocks demonstrates a test of the gravitational redshift of ${\alpha = (1.4 \pm 9.1)\times 10^{-5}}$ \cite{takamoto-2020}, not far behind the current best constraint (see also Ref.~\cite{liu-2024} for future prospects). Aside from testing LPI, atomic clocks underlie the field of relativistic geodesy as they can probe the geopotential at the sub-centimeter scale, see e.g. Refs.~\cite{zheng-2023, mcgrew-2018, delva-2019, bothwell-2022}.

The orders of magnitude mentioned in the above paragraphs will serve as benchmarks when we discuss redshift-based tests of the chameleon model, namely in Secs.~\ref{subsec:optimal-constraints} and \ref{sec:more-realistic-experiments}. Specifically, we will assess how these precision levels translate into constraints in the parameter space of the model.

\subsection{Principle of the experiment}
\label{subsec:experiment-principle}

Suppose we have two clocks, placed at different spatial positions, whose frequencies can be compared. It follows from Sec.~\ref{sec:gravitational-redshift-stt} that, for the scalar field $\phi$ to have an imprint on the measured redshift, the clocks must be immersed in different scalar field background values $\phi_1$ and $\phi_2$. At first sight, this can be achieved by placing the clocks in boxes filled with materials of different densities $\rho_1$ and $\rho_2$. Specifying to the chameleon model, it can be assumed that the scalar field will reach the value that minimizes its effective potential deep inside the boxes, provided they are large enough. This idea is developed in greater details in Appendix~\ref{app:thought-experiments} and is notably illustrated in Fig.~\ref{fig:gedankenexperiment}. In particular, it is shown there that the scalar contribution to the total redshift, denoted by $z_{\phi}$, can be approximated as
\begin{equation}
z_{\phi} \simeq \frac{\beta}{M_{\mathrm{Pl}}} (\phi_2 - \phi_1) \, .
\label{eqn:scalar-redshift}
\end{equation}

Now, we have not yet explained how this kind of experiment could be translated into constraints on the chameleon model. As an example, here is a proposal of a well-posed experiment:
\begin{enumerate}
\item We start with the two boxes filled with the same higher density material $\rho_1$. At first, there is no reason for the clocks to be synchronized\footnote{Here, we refer to frequency synchronization, also known as \textit{syntonization}.} as they could be at slightly different altitudes within the geopotential,\footnote{As a side note, nowadays we are able to resolve the gravitational potential of the Earth at the millimeter scale \cite{bothwell-2022}.} so we adjust their relative height so that $z \equiv 0$.
\item Then using pumps, we replace $\rho_1$ by $\rho_2$ (with ${\rho_2 \ll \rho_1}$) in one of the two boxes.
\item The frequency shift between the two clocks is measured again, without moving the boxes. In pure GR, the removed mass from the box affects the redshift through its Newtonian potential, which can readily be estimated. In scalar-tensor gravity, one has to further take into account the scalar field contribution $z_{\phi} \propto \Delta_{12} \phi$. The measured redshift together with its uncertainty can be used to put upper bounds on $| z_{\phi} |$, which in turn constrains the underlying scalar-tensor model.
\end{enumerate}
We stress the importance of precisely defining a protocol \textemdash \ altitudes, for instance, cannot be assumed to be known. Note that this protocol shares some similarities with neutron interferometry experiments in the sense that they also exploit the fact that the chameleon field takes different values in media of different densities, see e.g. Ref.~\cite{fischer-2024}.

\subsection{Optimal constraints}
\label{subsec:optimal-constraints}

\begin{table}[t]
    \centering
    \begin{ruledtabular}
    \begin{tabular}{ccc}
        Material designation & Density ($\mathrm{kg / m^3}$) & Density ($\mathrm{eV}^4$) \\\midrule
        Lead & $11.4 \times 10^3$ & $4.9 \times 10^{19}$ \\
        Water & $10^3$ & $4.3 \times 10^{18}$ \\
        Air & $1.225$ & $5.3 \times 10^{15}$ \\
        UHV & $10^{-10}$ & $4.3 \times 10^{5}$ \\
        XHV & $10^{-15}$ & $4.3$ \\
        IPM & $10^{-20}$ & $4.3 \times 10^{-5}$
    \end{tabular}
    \end{ruledtabular}
    \caption{Typical materials together with their densities (in $\mathrm{kg / m^3}$ and in $\mathrm{eV}^4$) considered in Sec.~\ref{subsec:gedankenexperiment}. `UHV' and `XHV' stand for \textit{ultra-high vacuum} and \textit{extremely-high vacuum}, and can be produced in the laboratory using sophisticated vacuum chambers. `IPM' stands for \textit{interplanetary medium} and represents the thinly scattered matter that exists between the planets and other large bodies of the Solar system. To put things into perspective, the density at the geostationary altitude is roughly $\sim O (10^{-19} \, \mathrm{kg / m^3})$.}
    \label{tab:material-densities}
\end{table}

\begin{figure*}[t]
\includegraphics[width=\textwidth]{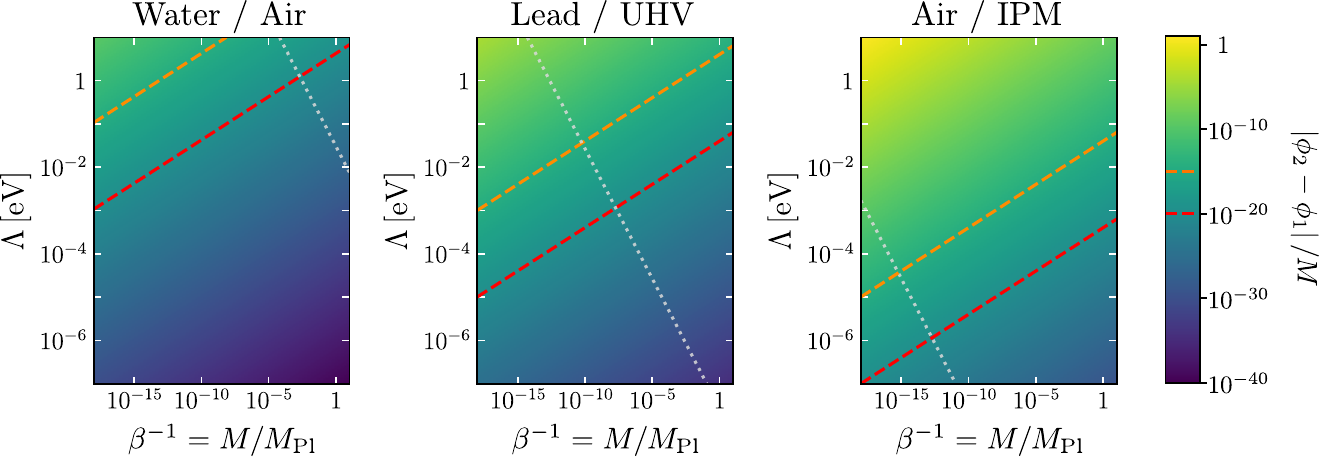}
\caption{Expected redshift from the chameleon field contribution Eq.~(\ref{eqn:scalar-contribution1}) for different pairs of materials. The chosen densities are reported in Table~\ref{tab:material-densities}. `UHV' stands for \textit{ultra-high vacuum} and corresponds to the vacuum level reachable in vacuum cavities while `IPM' stands for \textit{interplanetary medium}. The orange and red dashed lines correspond to iso-redshifts at levels $\varepsilon_{\mathrm{rel}} = 10^{-15}$ and $\varepsilon_{\mathrm{rel}} =10^{-20}$ respectively. The gray dotted line corresponds to $\lambda_c(\rho_2) = 1 \, \mathrm{m}$, where $\rho_2$ refers to the density of the less dense material of each pair: the Compton wavelength is larger (resp. smaller) than 1 meter above (resp. below) this line. The x-axis is $M / M_{\mathrm{Pl}} \equiv 1/\beta$ to be in line with the exclusion plots found in the literature, see e.g. Refs.~\cite{burrage-review-2018, chameleon-constraints-plot, fischer-review-2024}.}
\label{fig:thought-experiment-constraints}
\end{figure*}

Given the above, we shall approximate the scalar contribution to the total redshift by Eq.~(\ref{eqn:scalar-redshift}). We consider three pairs of `materials' to fill the boxes: (Water / Air), (Lead / UHV) and (Air / IPM). Here `UHV' stands for \textit{ultra-high vacuum} and corresponds to a vacuum level reachable in vacuum chambers, while `IPM' stands for \textit{interplanetary medium}~\cite{ipm-1967}. The associated densities are reported in Table~\ref{tab:material-densities}, in SI units ($\mathrm{kg / m^3}$) and in natural units ($\mathrm{eV}^4$).

In Fig.~\ref{fig:thought-experiment-constraints}, we represent the scalar field contribution to the redshift, $z_{\phi}$, in the $(\beta^{-1} , \, \Lambda)$-plane\footnote{Most references in the literature do indeed use $M = M_{\mathrm{Pl}} / \beta$ in place of $\beta$ in their exclusion plots, see e.g. Refs.~\cite{burrage-review-2018, chameleon-constraints-plot}.\label{fn:beta-M}} ($n\!=\!1$) for the three aforementioned pairs of materials. The bounds on $(\beta, \, \Lambda)$ are chosen large enough for the redshift to cover many orders of magnitudes, ranging from $\sim 1$ to $10^{-40}$. To put Sec.~\ref{subsec:atomic-clocks} into perspective, we depict by orange and red dashed lines the iso-redshift levels at $\varepsilon_{\mathrm{rel}}\!=\!10^{-15}$ and $\varepsilon_{\mathrm{rel}}\!=\!10^{-20}$ respectively, which correspond to levels that are achievable (or close to being achieved) given existing technologies. As it can be guessed from Eq.~(\ref{eqn:scalar-contribution1}), these iso-levels map to straight lines in the parameter space with $\log$-scaled axes. Unsurprisingly, the more accurate the clocks, the smaller the measurable redshift and thus the tighter the potential constraints on the chameleon parameter space. Additionally, it is worth noting that high density materials (water, lead) on the one hand, and low density materials (UHV, IPM) on the other hand, do not play symmetrical roles at all. Because of the dependence $\phi_{\min} \propto \rho^{-1/(n+1)}$, the lower-density material has more weight on the redshift. In plain language, lowering $\rho_2$ by one order of magnitude at fixed $\rho_1$ results in an increase of the redshift much greater than if we were to increase $\rho_1$ by one order of magnitude at fixed $\rho_2$. As a matter of fact, replacing the (Air / IPM) pair by (Lead / IPM) would not have any visible effect on the right panel of Fig.~\ref{fig:thought-experiment-constraints}.\footnote{$1-[\phi_{\min}(\rho_{\textsc{ipm}}) - \phi_{\min}(\rho_{\text{air}})] / [\phi_{\min}(\rho_{\textsc{ipm}}) - \phi_{\min}(\rho_{\text{lead}})] \leq 10^{-10}$.\label{fn:odg-computation}} For this reason, moving one clock from the Earth surface (where the ambient density is that of the air) to the underground [where the ambient density is in $O(10^3\,\mathrm{kg/m^3})$] would result in a totally negligible gain.\footref{fn:odg-computation}

\subsection{Challenges}
\label{subsec:challenges}

To a certain extent, these order-of-magnitude forecasts justify the present study since it appears that we can gain at least several orders of magnitude with respect to current constraints from laboratory experiments (see Fig.~4 from Ref.~\cite{chameleon-constraints-plot}). They can be considered as \textit{optimal} constraints because for any two densities $(\rho_1 , \, \rho_2)$, $|\phi_{\min}(\rho_2) - \phi_{\min}(\rho_1)|$ represents an upper bound for $\Delta_{12} \phi$. Overall, the potential constraints outlined in Fig.~\ref{fig:thought-experiment-constraints} remain overly optimistic for several reasons:
\begin{enumerate}
    \item They rely on the best atomic clocks ever built, which may not be well-suited for the experimental design one ends up choosing. This is all the more true if one thinks of space-borne experiments;
    \item We have assumed vacuum levels hardly reachable on Earth (especially the IPM density, see Table~\ref{tab:material-densities});
    \item We have assumed that the boxes were large enough in size for the scalar field to reach $\phi_{\min}$ at their center;
    \item We have assumed that the atomic clocks themselves do not perturb the scalar field profile inside the box, which is not realistic.
\end{enumerate}
The main goal of the remainder of this study is to take these points into account, and see whether some experimental design could realistically produce competitive constraints on the chameleon model. Points 1 and 4 are addressed in Sec.~\ref{subsec:space-experiments} and Sec.~\ref{subsec:laboratory-experiments} respectively, partly through numerical simulations. Point 3 is discussed in Appendix~\ref{subsec:constraints-finite-size-boxes}, and Fig.~\ref{fig:redshift-constraints-finite-boxes} revises the bounds shown in Fig.~\ref{fig:thought-experiment-constraints} by assuming meter radius boxes. The remainder of the present section is devoted to point 2 and looks at the consequences of approaching a perfect vacuum on the chameleon field.

Let us first comment the limit $\rho_2 \to 0$. Assuming that the chameleon field is indeed able to track the minimum of its effective potential, the scalar contribution to the redshift tends to infinity. In the face of this rather unphysical outcome, we have to take a closer look at the various assumptions that led to it.

First, no vacuum is truly perfect, thus the limit $\rho_2 \to 0$ should be replaced by $\rho_2 \to \rho_* > 0$. In the laboratory, vacuum is primarily measured by its absolute pressure, which can be translated into a density provided that other parameters (such as temperature or chemical composition) have been determined. Vacuum tubes typically reach $\sim 10^{8}$ particles per $\mathrm{cm}^3$, while cryopumped MBE\footnote{Molecular-beam epitaxy.} chambers can go down to densities as low as $\sim 10^5$ particles per $\mathrm{cm}^3$.\footnote{Assuming air with an average molar mass of $29 \, \mathrm{g / mol}$, this corresponds to densities of $5 \times 10^{-12} \, \mathrm{kg / m^3}$ for the vacuum tube and $5 \times 10^{-15} \, \mathrm{kg / m^3}$ for the cryopumped MBE chamber. This is in line with Ref.~\cite{burrage-atom-interferometry-2015} which assumes a density of $10^{-14} \, \mathrm{kg / m^3}$ inside a vacuum chamber.} Outer space gets even closer to `true' vacuum. Far enough from the Earth, at the altitude of geostationary satellites, the density of residual atmosphere neighbors $4 \times 10^{-19} \, \mathrm{kg / m^3}$. Density keeps decreasing as we go to interplanetary space ($\sim 11$ molecules per $\mathrm{cm^3}$), interstellar space ($\sim 1$ particle per $\mathrm{cm}^3$), and eventually intergalactic space ($\sim 10^{-6}$ particle per $\mathrm{cm}^3$) \cite{modern-vacuum-physics, ipm-1967}. The latter is the closest physical approximation of a perfect vacuum, with a density of $\sim 10^{-27} \, \mathrm{kg / m^3}$ assuming particles the mass of hydrogen. Ultimately, even if every particle of matter could somehow be removed from a given volume, quantum fluctuations ensure that the energy it contains is never quite zero, and so the chameleon does not diverge to $+ \infty$.


Two remarks have to be made regarding the above:
\begin{enumerate}
	\item Speaking of matter density on inter-galactic scales, the background value of the scalar field should match that predicted by its cosmological evolution. The latter can actually be smaller than $\phi_{\min} (\rho_{\mathrm{vac}})$, see e.g. the enlightening discussion in Ref.~\cite{hees-2012}, Sec.~\textsc{iv\,a}.
	\item Considering such rarefied environments (e.g. few thousands of particles per cubic meter) raises the question of the legitimacy of \textit{averaging} the density. Loosely speaking, does the chameleon field `perceive' a collection of isolated $N$ particles in the same way as a homogeneous medium? To our knowledge, the only work that examine this problem is Ref.~\cite{mota-shaw-2006}. There, the authors find, on the basis of analytical approximations, that the macroscopic Compton wavelength $\langle \lambda_{\phi} \rangle$ of the chameleon inside a screened body that is itself made of individual particles is
	\begin{equation*}
		\langle \lambda_{\phi} \rangle = \mathrm{max} \Bigl( m_{\phi}^{-1}\bigl(\langle \rho \rangle\bigr) , \, m_{\mathrm{crit}}^{-1} \Bigr) \, ,
	\end{equation*}
	where $\langle \rho \rangle$ denotes the average density of the body at stake while $m_{\mathrm{crit}}$ is a quantity depending solely on its microscopic properties and $n$. However, no insight is provided regarding the mean value of the chameleon field. This question can be investigated in more details numerically with \textit{femtoscope}~\cite{mypaper-nonlinear-kg-fem-2022} by using a representative volume element as the simulation box, with periodic boundary condition. This is described in the companion article~\cite{future-work}.
\end{enumerate}

Second, we have to verify that the expansion of the conformal factor around $1$ [Eq.~(\ref{eqn:conformal-factor-approximation})], which we have assumed in the derivation of the Newtonian limits of the redshift, holds. Since the maximum value of the chameleon field is given by $\phi_{\min} (\rho_{\min})$ [see Eq.~(\ref{eqn:chameleon-effective-min})], where $\rho_{\min}$ denotes the minimum density in the spatial region of interest, the condition $\phi \ll M_{\mathrm{Pl}} / \beta$ translates to
\begin{equation}
	\rho_{\min} \gg n \Lambda^{n+4} \left( \frac{\beta}{M_{\mathrm{Pl}}} \right)^{\! n} .
\label{eqn:approximation-omega-unity-chameleon}
\end{equation}
This condition is easily met for the $(\beta , \, \Lambda)$ ranges and material density considered in Fig.~\ref{fig:thought-experiment-constraints} \textemdash \ except in the top left corner of the parameter space ($\beta = 10^{18} , \, \Lambda = 10 \, \mathrm{eV}$) for which the rhs of Eq.~(\ref{eqn:approximation-omega-unity-chameleon}) reaches $\sim 10^{-20} \, \mathrm{kg / m^3}$, which corresponds to the IPM density. This zone of the parameter space is already well-constrained and thus not very relevant anyway. Finally, it should be reminded that this approximation was primarily performed for expanding the ratio of conformal factors $\Omega_{\mathrm{rec}} / \Omega_{\mathrm{em}}$ in Eq.~(\ref{eqn:gravitational-redshift-st}). Without this approximation and all other things being equal, the successive implications
\begin{equation*}
	\rho_2 \to 0 \implies \phi_{\min} (\rho_2) \to + \infty \implies \frac{\Omega_2}{\Omega_1} \to + \infty
\end{equation*}
remain true.

\section{Towards more realistic experimental designs}
\label{sec:more-realistic-experiments}

In the previous section, we imagined an idealized setup whereby atomic clocks are placed in different chameleon field backgrounds, which is achieved by adjusting the density of the medium in which they are immersed. The scalar field contribution $z_{\phi}$ to the total redshift between the two clocks is dominated by $\Delta_{12} (\beta \phi / M_{\mathrm{Pl}})$, while the Newtonian contribution $z_{N} = \Delta_{12} \Phi_N$ can readily be estimated (by calculation) since the mass content inside each box is assumed to be well-controlled. Using rough orders of magnitude, we evaluated the constraining power of such an experiment on the chameleon model.

The most controversial assumption we still have not discussed is the backreaction of the atomic clocks themselves on the scalar field profile. So far, we considered that the clocks were somehow `transparent' to the gravitational fields (scalar and metric in the Einstein frame), in the sense that the former would not significantly perturb background values of the latter. In GR, this is most likely true as the geopotential is overwhelmingly dominant over, say, laboratory-scale objects \textemdash \ this is because gravity is mediated by a \textit{massless} spin-two particle. In chameleon gravity however, the nonlinearity and mass-changing properties of the scalar field mean that the atomic clocks can be screened in the setup described above. This issue is of the utmost importance as it is an experiment killer: in this scenario, the interior of the clocks, where atoms are being `interrogated', becomes completely decoupled from the exterior, and therefore insensitive to the actual material filling the rest of the box. In that case, $z_{\phi}$ is expected to be essentially zero, meaning that the experiment cannot probe for the chameleon field.

In this section, we go a step further by taking these considerations into account. First, we examine more closely the implications of adding macroscopic atomic clocks to the model. We propose a more realistic redshift experiment in the laboratory that could be relevant for probing chameleons very strongly coupled to matter ($\beta \gtrsim 10^{5}$). Secondly, we revive the idea of space-based experiments as they could be sensitive to chameleons with gravitational strength coupling ($\beta \lesssim 10$).

\subsection{In the laboratory [very strong coupling]}
\label{subsec:laboratory-experiments}

Following on from Sec.~\ref{subsec:optimal-constraints}, we study what happens to the forecasts outlined in Figs.~\ref{fig:thought-experiment-constraints} and \ref{fig:redshift-constraints-finite-boxes} for chameleons strongly coupled to matter.

\subsubsection{Why the gedankenexperiment cannot be implemented}
\label{subsubsec:gedankenexperiment-fails}

An atomic clock relies on the interaction between two electron states in a given atom and some electromagnetic radiation. A group of atoms (e.g. cesium-133, rubidium-87 or strontium-87) is prepared in one energy state before being subjected to some monochromatic electromagnetic radiation, whose frequency is adjusted to match the targeted transition between the two energy states. Achieving this usually requires a whole apparatus that is rather bulky \textemdash \ see e.g. Fig.~1 from Ref.~\cite{riken-sr-clock-2021} showing Sr-based optical clocks at RIKEN laboratory. The most precise clocks are therefore quite large objects in the laboratory,\footnote{Early on, they even used to be the size of an entire room!} where the meter is a good characteristic length scale. Nonetheless, the past two decades have witnessed the development of chip-scale atomic clocks, a few centimeters in size and demonstrating a fractional frequency uncertainty of $2 \times 10^{-11}$ (see Ref.~\cite{csat-2019} and its supplementary material Table S1).

\begin{figure}[t]
\includegraphics[width=\linewidth]{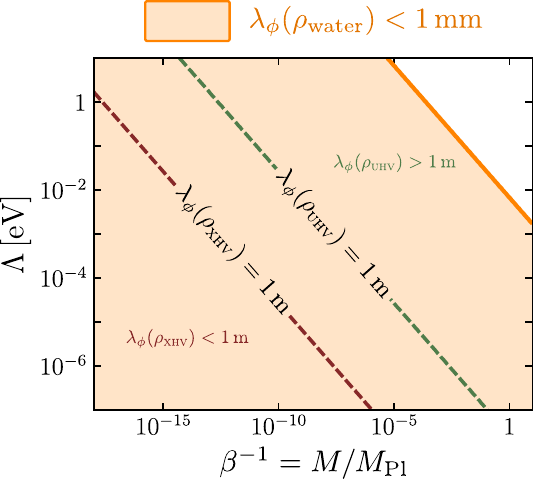}
\caption{The problem of Compton wavelengths. In the chameleon parameter space $(M, \, \Lambda , \,  n \! = \! 1)$,\footref{fn:beta-M} the two dashed lines represent the set of parameters that result in a one-meter Compton wavelength in the UHV and XHV vacua [see Eq.~(\ref{eqn:chameleon-compton-wavelength-bis}) and Table~\ref{tab:material-densities}]. The orange shaded area maps to sub-millimeter Compton wavelength in water \textemdash \ $\rho_{\mathrm{water}} = 10^3 \, \mathrm{kg / m^3}$ is representative of the typical density of materials found in the laboratory (including that of atomic clocks).}
\label{fig:materials-compton-wavelength}
\end{figure}

However small the actual clocks used in our \textit{gedankenexperiment} (see Appendix~\ref{subsec:gedankenexperiment}), they involve materials that are just too dense and too thick for it to be viable. Without going into too much details regarding the way an atomic clock is put together, it is conservative to assume that the average density of the apparatus is of the order of $\rho_{\mathrm{water}} = 10^3 \, \mathrm{kg / m^3}$, with walls of thickness greater than $1 \, \mathrm{mm}$ (even for the smaller chip-scale atomic clocks). In that respect, Fig.~\ref{fig:materials-compton-wavelength} provides insights into the various Compton wavelengths of the chameleon field involved in this experimental setup for the parameters $(M , \, \Lambda , \, n \! = \! 1)$. First, we saw with Fig.~\ref{fig:thought-experiment-constraints} that the lower density material has to be such that $\rho_2 \lesssim \rho_{\textsc{uhv}}$ for the experiment to yield interesting forecasts constraint-wise. At the same time, the constraint of the scalar field reaching $\phi_{\min}(\rho_i)$, $i \in \{1 , \, 2\}$ in finite-size boxes ($\sim 1 \, \mathrm{m}$) considerably restricts the region of the parameter space that can actually be probed. The two dashed lines in Fig.~\ref{fig:materials-compton-wavelength} represent the set of parameters that result in a one-meter Compton wavelength for the UHV and XHV vacua, while the orange area maps to $\lambda_{\phi} (\rho_{\mathrm{water}}) < 1 \, \mathrm{mm}$. An admissible region of the parameter space for the experiment to work would have to satisfy:
\begin{enumerate}
	\item yield a Compton wavelength in the clock's walls greater than their thickness, i.e. outside the orange shaded area;
	\item ensure that there is enough space in the box for the field to reach the value that minimizes the effective potential, i.e. below the dashed lines.
\end{enumerate}
Unfortunately, the intersection of these two regions is empty. In the region where condition 2 holds, the clock is expected to be deeply screened, thereby jeopardizing the whole concept of the thought experiment. The lessons drawn from this first experimental concept will nevertheless prove to be useful for the following.

\subsubsection{Alternative experimental design}
\label{subsubsec:alternative-experimental-design}


There may be nonetheless ways to benefit from atomic clocks, if one agrees to modify the experimental setup initially envisioned. In Sec.~\ref{subsec:optimal-constraints} and Appendix~\ref{subsec:gedankenexperiment}, we insisted on the need for high vacuum levels \textemdash \ the less dense, the better \textemdash \ for probing yet-unconstrained regions of the chameleon parameter space. As it turns out, atomic clocks also require such ultra-high vacuum environments to operate in optimal conditions. Indeed, this reduces background gas collisions in the atomic interrogation chamber (where the atoms interact with the electromagnetic radiation), the latter being detrimental to accuracy. This is true for both cesium / rubidium fountain clocks and for optical lattice clocks.

The idea is the following. We suppose that the science chamber is big enough for the chameleon field to reach $\phi_{\min}$ where the interrogated atoms sit. In order to modulate the scalar field they perceive, we cannot just increase the density inside the chamber as the atomic clock cannot operate correctly but in vacuum. Instead, we could imagine shrinking the chamber's size in order to bring its walls closer to atoms. The walls being dense and screened, this would effectively lower the chameleon field the atoms experience.

There are several shortcomings in this picture. A redshift measurement is a relative comparison between two frequencies, so we would need two clocks as before. One way to single out $z_{\phi}$ would be (\textit{i})~to start with two identical clocks with a `large' vacuum chamber, (\textit{ii})~synchronize them by adjusting their relative height, and (\textit{iii})~somehow shrink one clock's vacuum chamber and see how this affects the redshift. Having moving parts in a vacuum chamber, however, seems unfeasible. Actually, this is not needed. Since the regime we are probing here corresponds to high couplings of the scalar field to matter, any macroscopic object with density $\sim 10^3 \, \mathrm{kg/m^3}$ will be screened inside the vacuum chamber. Therefore, it is sufficient to bring such an object close enough to the atoms being interrogated to significantly alter the chameleon field they experience, all other things remaining equal. Below the dashed line $\lambda_{\phi} (\rho_{\textsc{uhv}}) = 1 \, \mathrm{m}$ in Fig.~\ref{fig:materials-compton-wavelength}, even an aluminum foil (which has a thickness of $\sim 0.2 \, \mathrm{mm}$) would be comfortably more than enough to screen the field. The closer the foil is to the atoms, the better. The material of the foil to be chosen, as well as the minimum distance at which it can be placed without disturbing the measurement is beyond the scope of this article.

\subsubsection{Orders of magnitude obtained from numerical computations}
\label{subsubsec:odg-numerical}

\begin{figure*}[t]
\includegraphics[width=\linewidth]{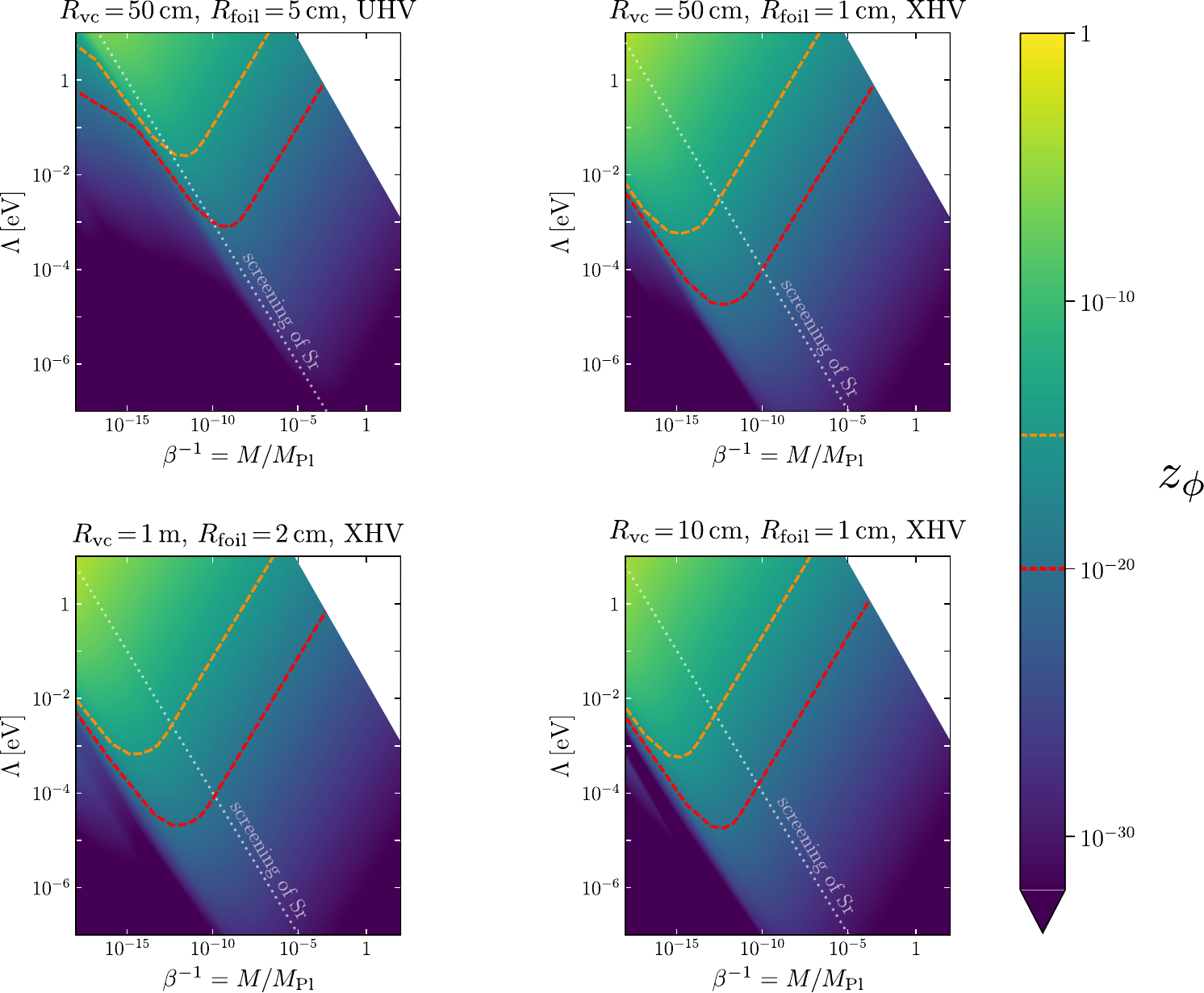}
\caption{Expected redshift from the chameleon field contribution $z_{\phi}$ for three different sets of parameters \textemdash \ the size of the vacuum chamber, the distance from the atoms to the foil and the vacuum density (UHV or XHV, see Table~\ref{tab:material-densities}) \textemdash \ in the chameleon parameter space ($n \! = \! 1$).\footref{fn:beta-M} The orange and red dashed lines correspond to the iso-redshift at $\varepsilon_{\mathrm{rel}} = 10^{-15}$ and $\varepsilon_{\mathrm{rel}} = 10^{-20}$ respectively. The white triangular mask in the top right corner of each panel correspond to a region of the parameter space where the foil is no longer screened (see Fig.~\ref{fig:materials-compton-wavelength}), which was not probed in the numerical computations. Below the diagonal dotted line, strontium nuclei are screened [see Eq.~(\ref{eqn:analytic-screening-criterion})].}
\label{fig:odg-constraints-foil}
\end{figure*}

In order to get an estimate of the part of the parameter space that can be probed with this idea of using a thin foil, we conduct 1D radial numerical computations with \textit{femtoscope} in three stages:
\begin{enumerate}
\item We compute the scalar field profile assuming a spherical vacuum chamber of radius $R_{\mathrm{vc}}$. Its walls are taken thick enough to be screened, so that the exterior environment has no influence whatsoever on the interior scalar field profile.
\item We then add the foil to the numerical domain, modeled as a spherical shell of density $\rho_{\mathrm{water}} = 10^3 \, \mathrm{kg / m^3}$, thickness $1 \, \mathrm{mm}$ and radius $R_{\mathrm{foil}}$, centered at the atoms' location. Note that the thickness parameter is not very relevant here since $\lambda_{\phi} (\rho_{\mathrm{water}})$ is smaller than the micrometer scale in the region $\lambda_{\phi} (\rho_{\textsc{uhv}}) < R_{\mathrm{vc}}$ probed here.
\item Finally, we estimate $z_{\phi}$ using the formula ${\beta |\phi_{\text{with foil}} - \phi_{\text{without foil}}| / M_{\mathrm{Pl}}}$.
\end{enumerate}
Of course, in reality putting a spherical shell around the atoms is absurd since it would block the electromagnetic radiation with which they have to interact. Nonetheless, this is deemed a good enough first approximation and allows for a relatively cheap numerical exploration of the full parameter space (since simulations are conducted in 1D).

The results of this simple study are presented in Fig.~\ref{fig:odg-constraints-foil}. As in Fig.~\ref{fig:thought-experiment-constraints}, we represent the scalar field contribution to the total redshift, $z_{\phi}$, for four different sets of the relevant parameters, namely $R_{\mathrm{vc}}$, $R_{\mathrm{foil}}$ and the density inside the vacuum chamber (UHV or XHV). The iso-redshift contours, at $10^{-15}$ (orange dashed line) and $10^{-20}$ (red dashed line), exhibit a typical `V' shape in the $(M, \, \Lambda)$-plane with log-scaled axes:
\begin{itemize}
	\item[--] \textit{left branch of the `V'} \textemdash \ In the lower left corner of each panel, the redshift suddenly drops to a very low level. This is due to fact that below a certain value of the $\alpha_{\scaleto{\mathrm{c}}{3pt}}$ parameter [Eq.~(\ref{eqn:alpha-c})], the Compton wavelength of the field in vacuum becomes smaller than $R_{\mathrm{foil}}$. As a result, the scalar field value at the atoms' location is the same with and without the foil, hence the vanishing redshift.
	\item[--] \textit{right branch of the `V'} \textemdash \ This is more or less the same behavior as the one exhibited in Figs.~\ref{fig:thought-experiment-constraints} and \ref{fig:redshift-constraints-finite-boxes}, although the interpretation is slightly different.  As we increase $\alpha_{\scaleto{\mathrm{c}}{3pt}}$ [Eq.~(\ref{eqn:alpha-c})], the Compton wavelength increases in the vacuum chamber. In either of the two configurations, the field has not enough space to reach $\phi_{\min} (\rho_{\mathrm{vac}})$ at its center, but takes nonetheless a higher value in the absence of the foil.
\end{itemize}
The \textit{sweet spot} is the bottom of the `V', that is when $\phi_{\text{without foil}} = \phi_{\min}(\rho_{\mathrm{vac}})$ but $\phi_{\text{with foil}} \ll \phi_{\min} (\rho_{\mathrm{vac}})$ \textemdash \ the foil playing its role in lowering the scalar field nearby the atoms. From the several sets of parameters tested (not all represented in Fig.~\ref{fig:odg-constraints-foil}), it appears that the best forecasts are obtained when the following three \textit{rules of thumb} are met: \textit{(i)} large vacuum chamber to give the field enough space to reach its highest value, (\textit{ii}) high vacuum level for maximizing the latter, (\textit{iii}) bringing the foil as close as possible to the atoms (without perturbing the measurement, which constitutes an open question).

Finally, it should be noted that the presence of the foil will also have an impact on the Newtonian potential, which in turn affects the total redshift through Eq.~(\ref{eqn:redshift-Newtonian-scalar-contributions}). The Newtonian potential at the center of a spherical shell of radius $R_{\mathrm{foil}}$, thickness $e \ll R_{\mathrm{foil}}$, and volumic density $\rho_{\mathrm{foil}}$ is simply
\begin{equation*}
	\Phi_{N , \, \mathrm{foil}} \simeq - 4 \pi e G \rho_{\mathrm{foil}} R_{\mathrm{foil}} \, .
\end{equation*}
For all four cases considered in Fig.~\ref{fig:odg-constraints-foil}, the contribution of $\Phi_{N , \, \mathrm{foil}}$ is many orders of magnitude below the sensitivity of the best atomic clocks, and one can thus ignore this term.

\subsubsection{Challenges}
\label{subsubsec:challenges}

We need to cast a critical eye on this setup idea. First, we have eluded the question of how to actually compare the scalar field value with and without the screened foil. One way to proceed, for instance, is to use a multiplexed optical lattice clock as in Refs.~\cite{zheng-2022, zheng-2023}, where two clouds of atoms are spatially separated in the same lattice and interrogated simultaneously by a shared clock laser and read-out in parallel. After performing a reference measurement of the gravitational redshift exactly as described in those references, the foil is added to surround only one of the two clouds of atoms and the measurement is repeated. The comparison of this second measurement against the reference one is then used to assert the consistency of the data within the chameleon model. Although such an experiment is more realistic than the first thought experiment exposed in Appendix.~\ref{subsec:gedankenexperiment}, it remains overly simplistic and might turn out to be unfeasible in practice. While addressing the corresponding technical issues is beyond the scope of this article, let us note that this proposal of putting a foil near some of the atoms is very close to the atom interferometry experiments described in Refs.~\cite{hamilton-2015, jaffe-2017, sabulsky-2019}.

Moreover, we have assumed throughout this discussion that the cloud of atoms does not perturb the chameleon field inside the vacuum chamber. We examine this hypothesis now. For atomic clocks based on optical transitions, what matters in redshift measurements is the value of the scalar field $\phi$ within the electron cloud \textemdash \ see Appendix~\ref{app:energy-transition} for a detailed explanation. On the one hand, matter density in atoms is mostly concentrated in the nucleus; thus deviations of the scalar field from its background value (i.e. in the absence of the atom) are mostly due to the presence of the nucleus (radial perturbation). On the other hand, the overlap of electronic wavefunctions peaks several Bohr radii away from the nucleus. Given the optical nuclear size of roughly $10^{-15} \, \mathrm{m}$ and the Bohr radius of roughly $10^{-10} \, \mathrm{m}$, one can expect the influence of the nucleus to be negligible on the scalar field value at the electron cloud. In order to assess this assumption, we perform a numerical experiment where we compare the scalar field value one \r{a}ngström away from the center of the nucleus, with and without the nucleus. These values are designated by $\phi_{\text{nucleus}}$ and $\phi_{\text{bg}}$ respectively. We further make several assumptions:
\begin{itemize}
\item[--] As in Sec.~\ref{subsubsec:odg-numerical}, the vacuum chamber is assumed to be spherical, with $R_{\mathrm{vc}}=5\,\mathrm{cm}$ and with walls thick enough to be screened.
\item[--] One atom of strontium, spherical likewise, is `put at the center' of the vacuum chamber. The nucleus is modeled by a sphere of radius $10^{-15} \, \mathrm{m}$ with a density of $2.5 \times 10^{17} \, \mathrm{kg/m^3}$. The electrons are assumed to be $10^{10} \, \mathrm{m}$ away from the nucleus and their mass density is neglected.
\item[--] The rest of the chamber is assumed to be a perfect vacuum. There is no issue with $\phi$ diverging towards $+ \infty$ because the walls of the chamber close off the setup.
\end{itemize}

\begin{figure}[t]
\includegraphics[width=\linewidth]{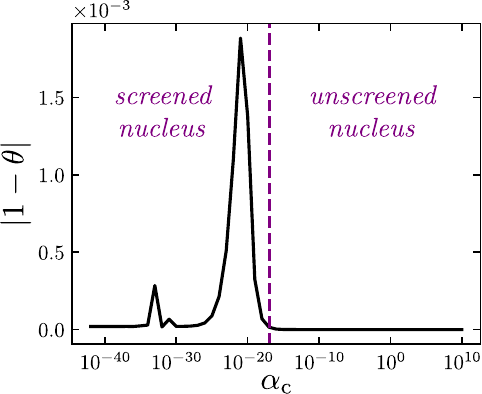}
\caption{Influence of the nucleus on the chameleon field at the electron cloud. Under the assumptions exposed in Sec.~\ref{subsubsec:challenges}, we numerically compute the chameleon field value $1\,\mathrm{fm}$ away from the nucleus center with and without the nucleus, denoted $\phi_{\text{nuc}}$ and $\phi_{\text{bg}}$ respectively. The ratio $\theta = \phi_{\text{nuc}} / \phi_{\text{bg}}$ is represented for several values of the dimensionless parameter $\alpha_{\scaleto{\mathrm{c}}{3pt}}$, ranging from $10^{-42}$ to $10^{10}$. The departure of this ratio from unity does not exceed $2 \times 10^{-3}$. The purple dashed line at $\alpha_{\scaleto{\mathrm{c}}{3pt}} \sim 10^{-17}$ is the threshold below which the atom's nucleus is screened.}
\label{fig:nucleus-screening}
\end{figure}

The results of this numerical experiment are displayed in Fig.~\ref{fig:nucleus-screening}. Specifically, we compute the ratio $\theta = \phi_{\text{nuc}} / \phi_{\text{bg}}$ for several values of the dimensionless parameter $\alpha_{\scaleto{\mathrm{c}}{3pt}}$ [see Eq.~(\ref{eqn:alpha-c})], ranging from $10^{-42}$ to $10^{10}$, which covers the whole parameter space displayed considered in this study (see Fig.~\ref{fig:alphas-spacecraft}). The purple dashed line corresponds to the threshold below which the nucleus becomes screened. The main highlight is that the departure of this ratio from unity does not exceed $2 \times 10^{-3}$, regardless of the regime. In other words, even when screened, the atom's nucleus has little to no influence on the chameleon field behavior at the electron cloud. Therefore, as far as atomic electron transitions are concerned and given the hypotheses we made, our experiment proposal appears to be unaffected by the screening of the nucleus.

To put things into perspective, we supplement each panel of Fig.~\ref{fig:odg-constraints-foil} with a diagonal dotted line, below which the nucleus of a strontium atom would be screened.\footnote{Note that this is in line with what is shown in Fig.~1 of Ref.~\cite{burrage-atom-interferometry-2015} for cesium and lithium atoms.} To this end, we employ the following quite common criterion, which states that a spherical body of density $\rho_i$ and radius $R_i$ immersed in a background chameleon field $\phi_{\mathrm{bg}}$ will be screened if
\begin{equation}
	\rho_i R_i^2 > 3 \frac{M_{\mathrm{Pl}}}{\beta} \phi_{\mathrm{bg}} \iff \alpha_{\scaleto{\mathrm{c}}{3pt}} < \frac{\bar{\rho}_i \bar{R}_i^2}{3 \bar{\phi}_{\mathrm{bg}}} \equiv \alpha_{\text{screened}} \, ,
\label{eqn:analytic-screening-criterion}
\end{equation}
where $\alpha_{\scaleto{\mathrm{c}}{3pt}}$ is the dimensionless parameter given by Eq.~(\ref{eqn:alpha-c}) and the bar notation is used to denote the associated dimensionless quantities. We evaluated the latter criterion in our numerical computations for the strontium-87 nucleus, although $\alpha_{\text{screened}}$ is hardly dependent on the specific isotope under consideration. Note that this analytical criterion yields comparable values to the numerical study conducted to produce Fig.~\ref{fig:nucleus-screening}.

\subsection{In space [gravitational strength coupling]}
\label{subsec:space-experiments}

The above considerations showed that laboratory-based redshift experiments are at best sensitive to very strongly coupled scalar fields. Specifically, they are completely blind to the region $\beta \lesssim 10^3$, which turns out to the least constrained by experiments~\cite{burrage-review-2018, chameleon-constraints-plot}. Given the above, we identify two main reasons for these limitations. On the one hand, we saw with the orders of magnitude laid out in Fig.~\ref{fig:thought-experiment-constraints} that for $\Lambda \lesssim 10^{-3} \, \mathrm{eV}$, the range $\beta \in [10^{-1} , \, 10^{5}]$ is only accessible in very low density environments. Yet, current vacuum technology has its limits. For instance, it does not allow us to reach the density levels found in the interplanetary medium. Nevertheless, laboratory experiments come with the constraint of \textit{size}. If we want to dictate the chameleon field value in two nearby regions of space, it has to be very dynamical and to closely follow density variations \textemdash \ this requires the field to be strongly coupled. The downside is that most laboratory objects end up being screened, including the atomic clocks themselves.

\subsubsection{On the screening of satellites}
\label{subsubsec:screening-satellites}

Going to space could precisely resolve these two limitations at once. Take a satellite in orbit around the Earth with an onboard atomic clock. Depending on its altitude, the background chameleon field in which it is immersed can be very high \textemdash \ see notably Ref.~\cite{mypaper-geodesy-2024}. For instance, at  geostationary altitude, the density is close to the `IPM' value tabulated in Table~\ref{tab:material-densities} and the scalar neighbors $\phi_{\min} (\rho_{\textsc{ipm}})$ (see e.g. Fig.~13 from Ref.~\cite{mypaper-nonlinear-kg-fem-2022}).

The sine qua non condition for hoping to use atomic clocks in space for constraining the chameleon model is that the spacecraft must be unscreened. Otherwise, the onboard clock will not see $\phi \sim \phi_{\min} (\rho_{\textsc{ipm}})$ but rather $\phi \sim 0$, and will therefore experience time as in GR. The assessment of whether a spacecraft orbiting the Earth is screened has been discussed several times in the literature \textemdash \ see e.g. Refs.~\cite{chameleon-KhouryWeltmanPRD, mota-shaw-2007, mpb-2019}. The key point that is usually stressed is that objects which possess thin shells down on Earth may loose them when they are taken into space. This can be seen by referring to the approximate screening criterion~(\ref{eqn:analytic-screening-criterion}): the low density environment offered by space, together with the large distance with respect to the Earth's surface, result in a higher background value for the scalar field $\phi_{\mathrm{bg}}$, which in turn means that the criterion is less easily satisfied. In Ref.~\cite{mypaper-geodesy-2024}, we went beyond this qualitative criterion by computing the full chameleon field of the ${ \{ \text{Earth} + \text{satellite} \} }$ system (without atmosphere though) using \textit{femtoscope} for various satellite's density and size in LEO. This showed, in line with the aforementioned qualitative argument, that a spacecraft could be unscreened in relevant parts of the parameter space \textemdash \ namely for sufficiently large values of $\alpha_{\scaleto{\mathrm{c}}{3pt}}$ (which, at fixed $\Lambda$, means small enough values of $\beta$).

\begin{table*}[t]
    \centering
    \begin{ruledtabular}
    \begin{tabular}{l c c c c c}
        Satellite & Mass & \makecell{Equivalent \\ radius} & \makecell{Mean \\ density} & \makecell{$\alpha_{\mathrm{screened}}$ \\ \footnotesize$ \rho_{\mathrm{bg}} \!=\! 10^{-12} \, \mathrm{kg / m^3}$} &  \makecell{$\alpha_{\mathrm{screened}}$ \\ \footnotesize$ \rho_{\mathrm{bg}} \!=\! 10^{-20} \, \mathrm{kg / m^3}$} \\ \midrule
CubeSat & $ 1 \, \mathrm{kg}$ & $ 6.2 \, \mathrm{cm}$ & $10^{3} \, \mathrm{kg / m^3}$ & $2 \times 10^{-6}$ & $2 \times 10^{-10}$  \\
\textsc{microscope} & $330 \, \mathrm{kg}$ & $ 82 \, \mathrm{cm}$ & $1.4 \times 10^{2} \, \mathrm{kg / m^3}$ & $5 \times 10^{-5}$ & $5 \times 10^{-9}$  \\
Galileo & $675 \, \mathrm{kg}$ & $95 \, \mathrm{cm}$ & $1.9 \times 10^{2} \, \mathrm{kg / m^3}$ & $9 \times 10^{-5}$ & $9 \times 10^{-9}$ \\
HST & $1.1 \times 10^{4} \, \mathrm{kg}$ & $5.6 \, \mathrm{m}$ & $15 \, \mathrm{kg / m^3}$ & $3 \times 10^{-4}$ & $3 \times 10^{-8}$
    \end{tabular}
    \end{ruledtabular}
\caption{Screening of satellites in space. Determination of $\alpha_{\mathrm{screened}}$ for two typical background densities ${ \rho_{\mathrm{bg}} \in \{ 10^{-12} \, \mathrm{kg/m^3} , \, 10^{-20} \, \mathrm{kg / m^3} \} }$ via 1D radial simulations performed with \textit{femtoscope}. The equivalent radius is computed such that a sphere of that radius would have the same volume as the actual satellite at stake. `Galileo' designate a satellite of the GNSS constellation of the same name, and `HST' is the acronym of the Hubble Space Telescope. We set $L_0 = 1 \, \mathrm{m}$, $\rho_0 = 1 \, \mathrm{kg / m^3}$, $n=1$.}
\label{tab:satellites-alpha-screened}
\end{table*}

Here however, we are interested in going to higher altitudes, farther away from the Earth's atmosphere where the vacuum is more pristine. In order to check whether a satellite is screened or not at such high altitudes, we use \textit{femtoscope} to solve the Klein\textendash Gordon equation governing the chameleon field of a ball immersed in a background medium of density $\rho_{\mathrm{bg}} \in \{ 10^{-12} \, \mathrm{kg / m^3} , \, 10^{-20} \, \mathrm{kg / m^3} \}$. Spherical symmetry allows for rapid radial computations, with the correct asymptotic boundary condition enabled. Specifically, we look for the value of the dimensionless parameter $\alpha_{\mathrm{screened}}$ below which the ball is screened (by dichotomy). Table~\ref{tab:satellites-alpha-screened} compiles such values for several satellites which are characterized by their mass and dimensions. Whatever the actual shape of the satellite at stake, we model it as a ball of equivalent density. We consider two background densities: $\rho_{\mathrm{bg}} = 10^{-12} \, \mathrm{kg / m^3}$ corresponds to the density found at an altitude of roughly $400 \, \mathrm{km}$, while $\rho_{\mathrm{bg}} = 10^{-20} \, \mathrm{kg / m^3}$ is the density representative of the IPM. The data obeys the scaling relation $\alpha_{\mathrm{screened}} \approxprop \tilde{\rho} \tilde{R}^2$, as expected from Eq.~(\ref{eqn:analytic-screening-criterion}). Nonetheless, we lay emphasis that this short study, despite being numerical, remains a crude approximation: in reality the atmospheric density varies non-isotropically away from the satellite, and the proximity to Earth (which is deeply screened in the range of parameters considered in Table~\ref{tab:satellites-alpha-screened}) further complicates the picture. Fig.~9 from Ref.~\cite{mypaper-geodesy-2024} clearly shows that, depending on both the actual atmospheric model and the altitude, the scalar field value can be either greater or lower than the value that minimizes the effective potential.

\begin{figure}[t]
\includegraphics[width=\linewidth]{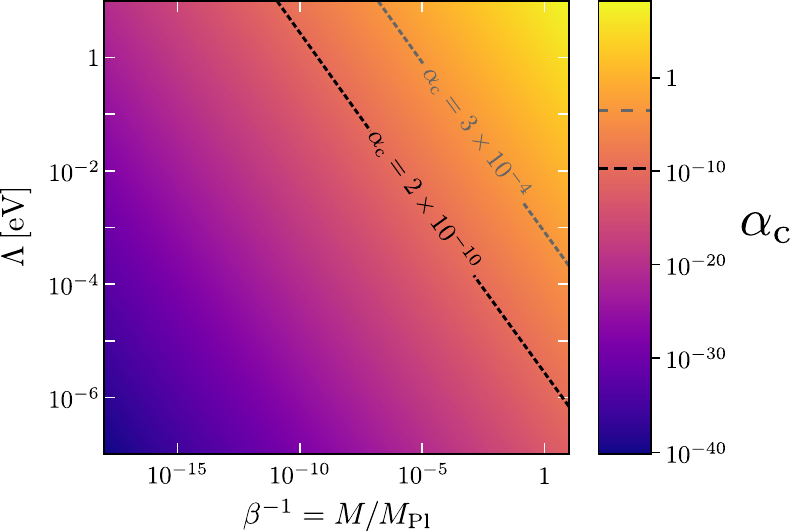}
\caption{Mapping from the chameleon parameter space $(M , \, \Lambda , \, n\!=\!1)$\footref{fn:beta-M} to the dimensionless $\alpha_{\scaleto{\mathrm{c}}{3pt}}$ parameter given by Eq.~(\ref{eqn:alpha-c}), which controls the behavior of the field up to a scaling. The iso lines $\alpha_{\scaleto{\mathrm{c}}{3pt}} \in \{2 \times 10^{-10} , \, 3 \times 10^{-4}\}$ are highlighted by the black and gray dashed lines respectively. They correspond to the minimum and maximum $\alpha_{\mathrm{screened}}$ values reported in Table~\ref{tab:satellites-alpha-screened}.}
\label{fig:alphas-spacecraft}
\end{figure}

Finally, it is worth noting that the case of an unscreened satellite orbiting in outer space is not too far off the the `clock in a box' picture that we used in our first thought experiment (see Sec.~\ref{subsec:experiment-principle} and Appendix~\ref{app:thought-experiments}), the box being made of space vacuum and having virtually no walls.

\subsubsection{Redshift measurements in space and ideas}
\label{subsubsec:space-ideas}

\begin{figure}[t]
\includegraphics[width=\linewidth]{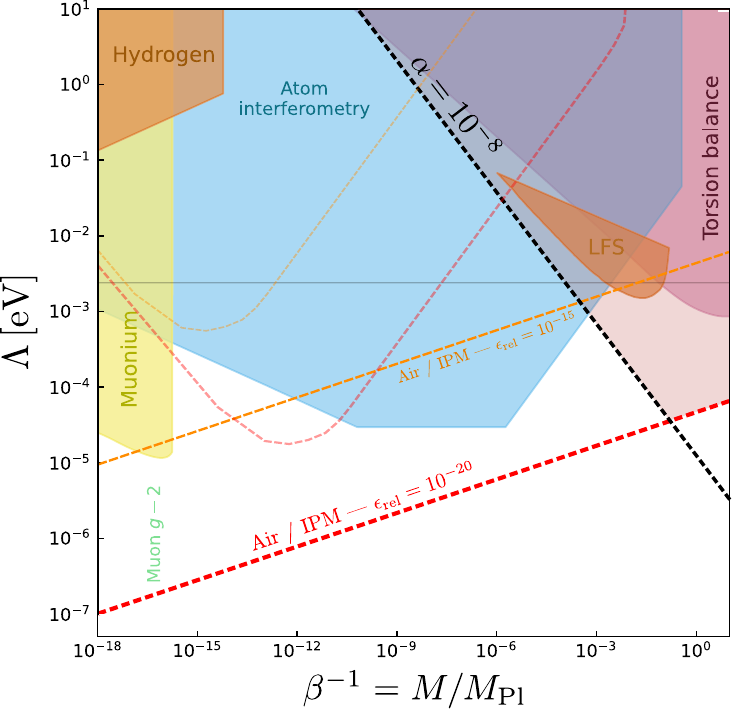}
\caption{Compilation of the forecasts coming from the redshift experiments \textemdash \ in the laboratory and in space \textemdash \ on the chameleon parameter space with state-of-the-art constraints (adapted from Fig.~4 of Ref.~\cite{chameleon-constraints-plot}).\footref{fn:beta-M} The laboratory constraints (`V'-shaped) are those from Fig.~\ref{fig:odg-constraints-foil} with parameters ${ \{ R_{\mathrm{vc}} \! = \! 50 \, \mathrm{cm} , \, R_{\mathrm{foil}} \! = \! 1 \, \mathrm{cm} , \, \text{XHV} \} }$ (top right panel). The region below the dotted gray line is where $^{87} \mathrm{Sr}$ nucleus starts to become screened and is thus, a priori, inaccessible to the atomic clock experiment described in Sec.~\ref{subsec:laboratory-experiments}. The shaded red area in the top right corner corresponds to the best possible constraints that could be set with a space-based experiment, regardless of the actual underlying mission concept. In this best case, highly optimistic scenario \textemdash \ small satellite beyond the geostationary altitude carrying a state-of-the-art atomic clock with a relative precision at $\varepsilon_{\mathrm{rel}} = 10^{-20}$ \textemdash \ one could access an unconstrained region of the chameleon parameter space, for gravitational strength couplings.}
\label{fig:constraint-plot-redshift}
\end{figure}

How does this translate into constraints on the chameleon model? While obtaining bounds from redshift measurements requires specifying an actual experiment, we can readily derive the best possible constraints by comparing $z_{\phi}$ to $\epsilon_{\mathrm{rel}}$. The scalar field redshift contribution $z_{\phi}$ is maximal for an unscreened satellite orbiting the Earth from a very high altitude (geostationary and beyond), where the ambient density is the lowest (more or less representative of the interplanetary medium, see Table~\ref{tab:material-densities}). In that case, an atomic clock onboard such a spacecraft would experience the very high value of the scalar field, which would be slightly lower but nonetheless close to $\phi_{\min} (\rho_{\textsc{ipm}})$. Comparing time as measured by this onboard clock against a ground-based reference one (or alternatively, one onboard a screened satellite orbiting at lower altitudes where the atmospheric density is several orders of magnitude higher than $\rho_{\textsc{ipm}}$) yields $z_{\phi} \sim \beta \phi_{\min}(\rho_{\textsc{ipm}}) / M_{\mathrm{Pl}}$. Such a value then constitutes a theoretical upper bound on $z_{\phi}$ in any realistic, well-defined experiment. In this perspective, Fig.~\ref{fig:constraint-plot-redshift} shows the associated best possible constraints in the chameleon parameter space for $ { \varepsilon_{\mathrm{rel}} \in \{ 10^{-15} , \, 10^{-20} \} } $ (red shaded area), together with the current constraints from other experiments (adapted from Fig.~4 of Ref.~\cite{chameleon-constraints-plot}). In particular, it is interesting to notice that space-based redshift measurements could open a new window for testing chameleons coupled to matter with gravitational strength \textemdash \ for $\beta \lesssim 10^{3}$.

Note that redshift measurements involving satellites mean that we can no longer assume that the emitter and the receiver are not moving in relation to each other. Consequently, an additional Doppler term from special relativity must be added to the redshift formula~(\ref{eqn:gravitational-redshift-st}). This Doppler effect can be larger than the `pure gravitational redshift' one \textemdash \ this is the case, for instance, for the ACES mission~\cite{aces-proposal-2019} where the overall rate of the clock onboard the ISS will be slower than a static ground one. This further complicates the data analysis process.

In this penultimate paragraph, we very briefly speculate on the actual feasibility of such space missions. First, we lay emphasis on the fact that clocks suitable for flying in space are not nearly as good as the best optical clocks engineered in the laboratory (given all the constraints inherent to space-flight). For instance, passive  hydrogen masers onboard Galileo satellites exhibit $O(10^{-14})$ stability levels at averaging times of $\sim 1000  \, \mathrm{s}$, which is roughly one order of magnitude better than the cesium clocks onboard \textsc{glonass} satellites. The ACES mission, to be launched in 2025, will put a cold atom cesium clock (PHARAO) in the International Space Station, targeting a $O(10^{-16})$ precision \cite{aces-proposal-2019}.

Given these orders of magnitude, it appears that space-borne atomic clocks are not mature enough yet to probe still unexplored regions of the chameleon parameter space \textemdash \ see Fig.~\ref{fig:constraint-plot-redshift}. Provided that they will continue to improve by a few orders of magnitude in the future, one could draw from past proposal for testing the gravitational redshift effect in space \cite{vessot-levine-1980, galileo1, galileo2, radioastron-2023}. In particular, highly elliptical orbits are good candidates for such tests in several respects. Regardless of the model being tested, an elliptic orbit induces a periodic modulation of the gravitational redshift~\cite{review-gps-2003}. In chameleon gravity, one could imagine having the science payload screened at the perigee, where the atmosphere is still thick enough, yet unscreened at the apogee, where density drops below $10^{-19} \, \mathrm{kg / m^3}$. In this regard, the RadioAstron space mission has the orbital characteristics that we seek, with a perigee at $500 \, \mathrm{km}$ above the Earth's surface, where the atmospheric density is roughly $10^{-12} \, \mathrm{kg / m^3}$, and very high apogee at an altitude of $350 \, 000 \, \mathrm{km}$, where density is that of the interplanetary medium. As reported in Table~\ref{tab:satellites-alpha-screened} and Fig.~\ref{fig:alphas-spacecraft}, there is a band in the parameter space for which a spacecraft would possess a thin shell at the perigee but would be unscreened at the apogee. This would leave a chameleon-modulated signal in the redshift data. Likewise, the FOCOS proposal~\cite{focos-proposal-2022}, with its highly eccentric orbit and state-of-the-art optical atomic clock onboard, could yield competitive constraints.

\section{Conclusion and perspectives}
\label{sec:conclusion}

\subsection{Summary of the study}

This article clarified, in theoretical terms, the predicted outcome of redshift measurements in the framework of scalar-tensor theories of gravity. Despite satisfying LPI, such theories can nonetheless be distinguished from GR in redshift experiments. This opens new venues for testing models that come with screening mechanisms \textemdash \ see Table~\ref{tab:tests-classification}.

Focusing on the chameleon model with positive exponent, we are able to single out the scalar field contribution to the total redshift (in the Newtonian limit), which beyond improving readability, allows us to imagine a first \textit{gedankenexperiment} involving atomic clocks and aimed at either detecting or constraining the model at stake. The orders of magnitude derived from this idealized yet well-defined experiment are translated into constraints in the parameter space of the chameleon, given the current state of the art in atomic clock technology. It appears that these `optimal' theoretical constraints are competitive with current bounds, which is why we go a step further and assess whether they hold in a more realistic scenario.

In the laboratory, modular atomic clocks could be sensitive to chameleons very strongly coupled to matter {($\beta \gg 10^5$)}. The experimental setup involving a foil placed around some of the atoms might actually be feasible, given its similarity in terms of apparatus with already-performed atom interferometry experiment. Furthermore, preliminary studies combined with theoretical arguments seems to indicate that such an experiment is not limited by the screening of atomic nuclei, allowing it to be sensitive to a large part of the chameleon parameter space. In space, the very low-density environment found in high-altitude orbits allows spacecraft to be unscreened in some regions of the parameter space \textemdash \ most notably for gravitational strength couplings. However, the clocks onboard satellites are currently not as accurate as their ground-based counterparts, mostly due to the constraints inherent to space-flight. We find that the current level of accuracy exhibited by the best clocks in space is still a few orders of magnitude too low to yield interesting constraints, although future space missions could make a difference~\cite{focos-proposal-2022}.

\subsection{Perspectives}

This paper focused on the very specific case of the chameleon model with exponent $n=1$. Because $\phi_{\min} (\rho) \propto \rho^{-1 / (n+1)}$, models with $n < 0$ do not exhibit the crucial property that $\phi_{\min} (\rho) \to + \infty$ as $\rho \to 0$, on which the whole idea developed in this paper holds. In the symmetron model, for which Eq.~(\ref{eqn:functions-chameleon}) has to replaced by
\begin{equation}
\Omega(\phi) = 1 + \frac{\phi^2}{2 M^2} \quad \text{and} \quad V(\phi) = -\frac{\mu^2}{2} \phi^2 + \frac{\lambda}{4} \phi^4 \, ,
\label{eqn:functions-symmetron}
\end{equation}
the scalar field has a non-zero vacuum expectation value in very low density environments, reading
\begin{equation}
	\phi_{\min} (\rho) = \pm \frac{\mu}{\sqrt{\lambda}} \sqrt{1 - \frac{\rho}{\mu^2 M^2}} \simeq \pm \frac{\mu}{\sqrt{\lambda}} \, ,
\end{equation}
which readily implies
\begin{equation}
\Omega(\phi_{\min}) \simeq 1 + \frac{\mu^2}{2 \lambda M^2} \, .
\end{equation}
In this model, the dimensionless parameter $\lambda$ can be very low (values considered in the literature go down to ${\lambda \lesssim 10^{-60}}$), while $\log_{10} \mu$ is generally taken between $-3$ and $+3$ and $\log_{10}(M / \mathrm{GeV})$ between $-10$ and $+20$. A rough order-of-magnitude computation seems to indicate that redshift experiments could yield interesting bounds on this model. The careful analysis of this question is left for future work.

Additionally, Sec.~\ref{subsec:challenges} raised the problem of the behavior of the scalar field in a diluted gas. Indeed, the validity of treating such low-density environment as homogeneous media has not been investigated thoroughly in the literature yet. Preliminary results indicate that, in the general case
\begin{equation}
\bigl\langle \phi(\mathbf{x}) \bigr\rangle \neq \phi_{\mathrm{min}} \bigl(\langle \rho(\mathbf{x}) \rangle\bigr) \, .
\end{equation}
This is ongoing work~\cite{future-work}.

\begin{acknowledgments}
The authors would like to thank Joël Bergé and Gilles Esposito-Farèse for their comments and insights. The authors would also like to thank the reviewers for their constructive comments, which have significantly improved the quality of this article.
\end{acknowledgments}

\appendix

\section{Additional derivations}
\label{app:additional-derivations}

\subsection{Jordan-frame fields equations}
\label{subsec:field-equations-jf}

The theory can be rewritten equivalently in the Jordan frame. To do so, we perform the well-known field redefinition $\phi \to \varphi$ by making use of three functions $F$, $H$ and $Z$, such that~\cite{gef-polarski-2001}
\begin{subequations}
\begin{align}
    F(\varphi) &= \Omega^{-2}(\phi) \, , \label{subeqn:F-function}\\[3pt]
    H(\varphi) &= \Omega^{-4}(\phi) \, V(\phi) \, , \label{subeqn:U-function}\\[3pt]
    \left( \frac{\mathrm{d} \phi}{\mathrm{d} \varphi} \right)^{\!2} &= \frac{Z(\varphi)}{F(\varphi)} + \frac{3}{2} M_{\mathrm{Pl}}^2 \left( \frac{\mathrm{d} \ln F}{\mathrm{d} \varphi} \right)^{\! 2} \, . \label{subeqn:Z-function}
\end{align}
\label{eqn:scalar-field-redefinition}%
\end{subequations}
Eqs.~(\ref{eqn:metric-eqn-ef}\textendash\ref{eqn:scalar-field-eqn-ef}) then become
\begin{align}
F(\varphi) \tilde{G}_{\mu \nu} &= \frac{1}{M_{\mathrm{Pl}}^2} \Bigl( \tilde{T}_{\mu \nu} + \tilde{T}_{\mu \nu}^{\scaleto{(\varphi)}{5pt}} \Bigr) \, , \label{eqn:metric-eqn-jf} \\[3pt]
Z(\varphi) \tilde{\Box}\varphi &= \frac{\mathrm{d} H}{\mathrm{d} \varphi}\!-\!\frac{M_{\mathrm{Pl}}^2}{2} \frac{\mathrm{d} F}{\mathrm{d} \varphi} \tilde{R} - \frac{1}{2} \frac{\mathrm{d} Z}{\mathrm{d} \varphi} \tilde{g}^{\alpha \beta} \partial_{\alpha} \varphi \, \partial_{\beta} \varphi . \label{eqn:scalar-field-eqn-jf}
\end{align}
In these equations, $\tilde{G}_{\mu \nu}$ is the Einstein tensor constructed from $\tilde{g}_{\mu \nu}$, $\tilde{T}_{\mu \nu}$ is the Jordan-frame stress-energy tensor of matter given by
\begin{equation}
\tilde{T}_{\mu \nu} = \frac{-2}{\sqrt{-\tilde{g}}} \frac{\delta S_{\mathrm{mat}}}{\delta \tilde{g}^{\mu \nu}} \, ,
\label{eqn:stress-energy-tensor-jf}
\end{equation}
and $\tilde{T}_{\mu \nu}^{\scaleto{(\varphi)}{5pt}}$ is a convenient notation for the scalar field contribution
\begin{equation}
\begin{split}
\tilde{T}_{\mu \nu}^{\scaleto{(\varphi)}{6pt}} = & \, Z(\varphi) \! \left[ \partial_{\mu} \varphi \, \partial_{\nu} \varphi - \frac{1}{2} \tilde{g}_{\mu \nu} \tilde{g}^{\alpha \beta} \partial_{\alpha} \varphi \, \partial_{\beta} \varphi \right] \\[3pt]
& - \tilde{g}_{\mu \nu} U(\varphi) + M_{\mathrm{Pl}}^2 \Bigl( \tilde{\nabla}_{ \!\mu} \tilde{\nabla}_{\!\nu} F - \tilde{g}_{\mu \nu} \tilde{\Box} F \Bigr) \, .
\end{split}
\label{eqn:stress-energy-tensor-scalar-jf}
\end{equation}
Note that we have further defined the d'Alembertian $\tilde{\Box} \equiv \tilde{g}^{\mu \nu} \tilde{\nabla}_{\! \mu} \tilde{\nabla}_{\! \nu}$, where $\tilde{\nabla}_{\! \mu}$ refers to the covariant derivative constructed from $\tilde{g}_{\mu \nu}$.

\subsection{Stress-energy tensors of matter in both frames}
\label{subsec:stress-energy-tensors-both-frames}

The definitions of the stress-energy tensor of matter in the Einstein frame~[Eq.~(\ref{eqn:stress-energy-tensor-ef})] and in the Jordan frame~[Eq.~(\ref{eqn:stress-energy-tensor-jf})] imply the following relations
\begin{equation}
\tilde{T}_{\mu \nu}= \Omega^{-2} T_{\mu \nu} \, , \quad \tilde{T}^{\mu \nu} = \Omega^{-6} T^{\mu \nu} \, , \quad \tilde{T} = \Omega^{-4} T \, .
\label{eqn:stress-energy-tensor-matter-relations}
\end{equation}
For a perfect fluid source of energy-momentum with 4-velocity $\tilde{u}^{\mu}$, $\tilde{\rho}$ and $\tilde{p}$ as rest-frame energy and momentum densities \textemdash \ in the Jordan frame \textemdash, we have
\begin{equation}
\begin{split}
\tilde{T}_{\mu \nu} &= (\tilde{\rho} + \tilde{p}) \tilde{u}_{\mu} \tilde{u}_{\nu} + \tilde{p} \tilde{g}_{\mu \nu} \\[3pt]
&= \Omega^2 \bigl[ (\tilde{\rho} + \tilde{p}) u_{\mu} u_{\nu}+ \tilde{p} g_{\mu \nu} \bigr] = \Omega^{-2} T_{\mu \nu} \, .
\end{split}
\label{eqn:stress-energy-tensors-perfect-fluid}
\end{equation}
Therefore, it makes sense to define ${(\rho , \, p) \equiv \Omega^4 (\tilde{\rho} , \, \tilde{p})}$ so that we recover the canonical form
\begin{equation}
T_{\mu \nu} = (\rho + p) u_{\mu} u_{\nu} + p g_{\mu \nu} \, .
\label{eqn:stress-energy-tensor-perfect-fluid-ef}
\end{equation}
Indeed, ${g_{\mu \nu} u^{\mu} u^{\nu} = -1 = \tilde{g}_{\mu \nu} \tilde{u}^{\mu} \tilde{u}^{\nu} }$, which readily implies ${\tilde{u}^{\mu} = \Omega u^{\mu}}$.

\subsection{Newtonian limit in the Jordan frame}
\label{subsec:jf-newtonian-limit}

In order to derive the Newtonian limit of Eqs.~(\ref{eqn:metric-eqn-jf}\textendash\ref{eqn:scalar-field-eqn-jf}), we proceed as in Sec.~\ref{subsec:field-equations-newtonian-limits} by expanding the Jordan-frame metric about the Minkowski metric as ${\tilde{g}_{\mu \nu} = \eta_{\mu \nu} + \tilde{h}_{\mu \nu}}$, with ${|\tilde{h}_{\mu \nu}| \ll 1}$. Again, gauge freedom allows us to put the metric in the form
\begin{equation}
d\tilde{s}^2 \equiv \tilde{g}_{\mu \nu} \mathrm{d}x^{\mu} \mathrm{d}x^{\nu} \! = -(1+ 2 \tilde{\Phi}) \mathrm{d}t^2 + \tilde{g}_{ij} \mathrm{d}x^i \mathrm{d}x^j ,
\label{eqn:newton-gauge-jf}
\end{equation}
featuring a potential $\tilde{\Phi}$, with $|\tilde{\Phi}| \ll 1$. We stress that the assumption $|\tilde{h}_{\mu \nu}| \ll 1$ is a priori not equivalent to $|h_{\mu \nu}| \ll 1$. Indeed, without making any approximation, it follows from Eq.~(\ref{eqn:weyl-transform}) that
\begin{equation*}
\tilde{h}_{\mu \nu} = \bigl( \Omega^2 - 1 \bigr) \eta_{\mu \nu} + \Omega^2 h_{\mu \nu} \, .
\end{equation*}
Therefore, we see that the consistency between the Newtonian limits in the Einstein frame [Eq.~(\ref{eqn:newton-gauge-ef})] and in the Jordan frame [Eq.~(\ref{eqn:newton-gauge-jf})] is ensured by condition~(\ref{eqn:conformal-factor-approximation}). The 00-component of Eq.~(\ref{eqn:metric-eqn-jf}) then becomes
\begin{equation}
M_{\mathrm{Pl}}^2 \Bigl[ 2 F(\varphi) \tilde{\Delta} \tilde{\Phi} + \tilde{\Delta} F \Bigr] = \tilde{\rho} - 2 U(\varphi) \, ,
\label{eqn:metric-eqn-newtonian-limit-jf}
\end{equation}
while the scalar field equation~(\ref{eqn:scalar-field-eqn-jf}) simplifies to
\begin{equation}
Z(\varphi)  \tilde{\Delta} \varphi = \frac{\mathrm{d} H}{\mathrm{d} \varphi} - \frac{1}{2} \frac{\mathrm{d} \ln F}{\mathrm{d} \varphi} \Bigl[ \tilde{\rho} + 4 H(\varphi) + 3 M_{\mathrm{Pl}}^2 \tilde{\Delta} F \Bigr] .
\end{equation}
Here, the Laplacian is formally defined as $\tilde{\Delta} \equiv \tilde{g}^{ij} \partial_i \partial_j$.

\subsection{Null geodesics}
\label{subsec:null-geodesics}

It is obvious that conformal transformations leave null \textit{curves} invariant, since ${\tilde{g}_{\mu \nu} \mathrm{d}x^{\mu} \mathrm{d} x^{\nu} = 0}$ is equivalent to ${g_{\mu \nu} \mathrm{d}x^{\mu} \mathrm{d} x^{\nu} = 0}$. One can actually show that they leave null \textit{geodesics} invariant as well \textemdash \ which is a stronger result \textemdash \ see e.g. Ref.~\cite{wald-1984}, Appendix G. Denoting ${k^{\mu} = \mathrm{d}x^{\mu} / \mathrm{d} \lambda}$ the tangent vector to a null geodesic of the Einstein-frame metric, affinely parameterized by $\lambda$, we have
\begin{equation}
k^{\alpha} \nabla_{\! \alpha} k^{\mu} = 0 \, .
\label{eqn:null-geodesic-ef}
\end{equation}
One can further define a new affine parameter $\tilde{\lambda}$, related to $\lambda$ via $\mathrm{d}\tilde{\lambda} = \Omega^2 \mathrm{d} \lambda$, so that
\begin{equation}
\tilde{k}^{\alpha} \tilde{\nabla}_{\! \alpha} \tilde{k}^{\mu} = 0 \, , \qquad \text{with} \quad \tilde{k}^{\mu} = \frac{\mathrm{d} x^{\mu}}{\mathrm{d} \tilde{\lambda}} = \Omega^{-2} k^{\mu} \, .
\label{eqn:null-geodesic-jf}
\end{equation}

\section{Derivation of the redshift expression}
\label{app:redshift-expression}

\begin{figure}[t]
\includegraphics[width=\linewidth]{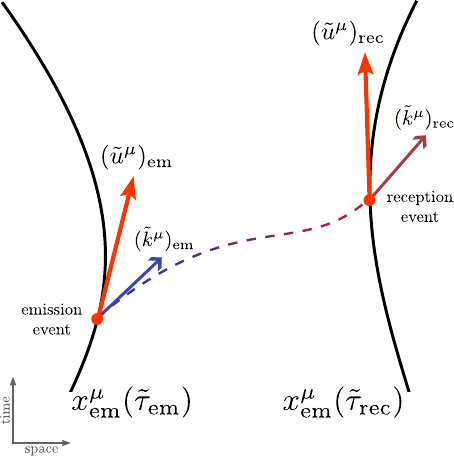}
\caption{Notations associated with the redshift definition on a spacetime diagram. The solid black line labeled $x^{\mu}_{\mathrm{em}}$ (resp. $x^{\mu}_{\mathrm{rec}}$) is the worldline of the emitter (resp. receiver), parametrized by the proper time $\tilde{\tau}_{\mathrm{em}}$ (resp. $\tilde{\tau}_{\mathrm{rec}}$) and with 4-velocity $(\tilde{u}^{\mu})_{\mathrm{em}}$ (resp. $(\tilde{u}^{\mu})_{\mathrm{rec}}$). $(\tilde{k}^{\mu})_{\mathrm{em}}$ (resp. $(\tilde{k}^{\mu})_{\mathrm{rec}}$) denotes the photon 4-wave-vector at the emission event (resp. reception event). The dashed line corresponds to the photon's null geodesic between the two events.}
\label{fig:spacetime-diagram-redshift}
\end{figure}

Here we derive the redshift expression in the framework of scalar-tensor theories given by the action~(\ref{eqn:action-ef}\textendash\ref{eqn:action-parts}). One observer, the \textit{emitter}, sends a photon to another observer, the \textit{receiver}. The gravitational redshift, denoted by $z$, is \textit{defined} as
\begin{equation}
    z \equiv \frac{E_{\mathrm{em}}}{E_{\mathrm{rec}}} - 1 \, ,
\label{eqn:redshift-definition}
\end{equation}
where $E_{\mathrm{em}}$ (resp. $E_{\mathrm{rec}}$) denotes the energy of the photon \textit{measured} by the emitter (resp. by the receiver). The \textit{physical} metric to be used in the subsequent calculations of these energies is the Jordan-frame metric $\tilde{g}_{\mu \nu}$. Indeed, it is the metric to which matter is universally coupled [see Eq.~(\ref{subeqn:matter-action})] and thus defines the lengths and times measured by material rods and clocks \textemdash \ see Refs.~\cite{hughes-1990, gef-2011}. Consequently, we have
\begin{equation}
    \frac{E_{\mathrm{em}}}{E_{\mathrm{rec}}} = \frac{ (\tilde{u}_{\mu} \tilde{k}^{\mu})_{\mathrm{em}}}{ (\tilde{u}_{\mu} \tilde{k}^{\mu})_{\mathrm{rec}}} \, ,
\end{equation}
where $(\tilde{u}^{\mu})_{\mathrm{em}}$ (resp. $(\tilde{u}^{\mu})_{\mathrm{rec}}$) denotes the 4-velocity of the emitter (resp. receiver) and $\tilde{k}^{\mu}$ represents the null tangent vector of a geodesic joining the emission and reception events (where the affine parameter is normalized so that $\tilde{k}^{\mu}$ coincides with the 4-wavevector). These quantities are defined in the Jordan frame, and normalized such that
\begin{equation}
    \tilde{g}_{\mu \nu} \tilde{u}^{\mu} \tilde{u}^{\nu} = -1 \qquad \text{and} \qquad \tilde{g}_{\mu \nu} \tilde{k}^{\mu} \tilde{k}^{\nu} = 0 \, .
\end{equation}
For the sake of clarity, notations are illustrated on a spacetime diagram in Fig.~\ref{fig:spacetime-diagram-redshift}.

Let us now make some additional assumptions:
\begin{enumerate}
    \item The spatial coordinates of the two observers, $(x^i)_{\mathrm{em}}$ and $(x^i)_{\mathrm{rec}}$, remain constant throughout the experiment.
    \item The metric $\tilde{g}_{\mu \nu}$ is stationary, meaning that it does not depend upon the $x^0$ coordinate, i.e. $\partial_0 \tilde{g}_{\mu \nu} = 0$. Consequently, the scalar field $\phi$ will also be assumed stationary [which has already been assumed in Eq.~(\ref{eqn:scalar-field-eqn-newtonian-limit-ef})].
\end{enumerate}
As a direct consequence of the first assumption, the relation $\tilde{g}_{\mu \nu} \tilde{u}^{\mu} \tilde{u}^{\nu} = -1$ leads to $\tilde{u}^0 = 1/\sqrt{-\tilde{g}_{00}}$ (since $\tilde{u}^i = 0$). Moreover, the fact that the metric is taken stationary implies that there exists a timelike Killing vector $\tilde{\boldsymbol{\xi}} = (1, \, 0, \, 0, \, 0)$ associated with the time translation symmetry. Denoting $\tilde{\lambda}$ the affine parameter of the photon geodesic, mathematical properties of Killing vectors (see e.g. Ref.~\cite{redshift-killing-vectors}) let us write
\begin{equation}
    \frac{\mathrm{d}}{\mathrm{d} \tilde{\lambda}} \Bigl(\tilde{\xi}_{\mu}\tilde{k}^{\mu}\Bigr) = 0 \implies \frac{\mathrm{d}}{\mathrm{d} \tilde{\lambda}} \bigl(\tilde{k}_0\bigr) = 0 \, .
\label{eqn:killing-vector-property-redshift}
\end{equation}
Thus, $\bigl(\tilde{k}_0\bigr)_{\!\mathrm{em}} = \bigl(\tilde{k}_0\bigr)_{\!\mathrm{rec}}$ and we eventually get
\begin{equation}
	1+z = \sqrt{\frac{(\tilde{g}_{00})_{\mathrm{rec}}}{(\tilde{g}_{00})_{\mathrm{em}}}} = \frac{\Omega_{\mathrm{rec}}}{\Omega_{\mathrm{em}}} \sqrt{\frac{(g_{00})_{\mathrm{rec}}}{(g_{00})_{\mathrm{em}}}} \, .
\label{eqn:gravitational-redshift-st-app}
\end{equation}

\section{Thought experiments}
\label{app:thought-experiments}

\subsection{A first gedankenexperiment}
\label{subsec:gedankenexperiment}

Here, we imagine a toy experiment to prove the point we made in Sec.~\ref{subsec:link-observable-quantities}, namely that it is possible to distinguish a scalar-tensor theory complying with LPI from GR by means of redshift measurements.

\begin{figure*}[t]
\includegraphics[width=\linewidth]{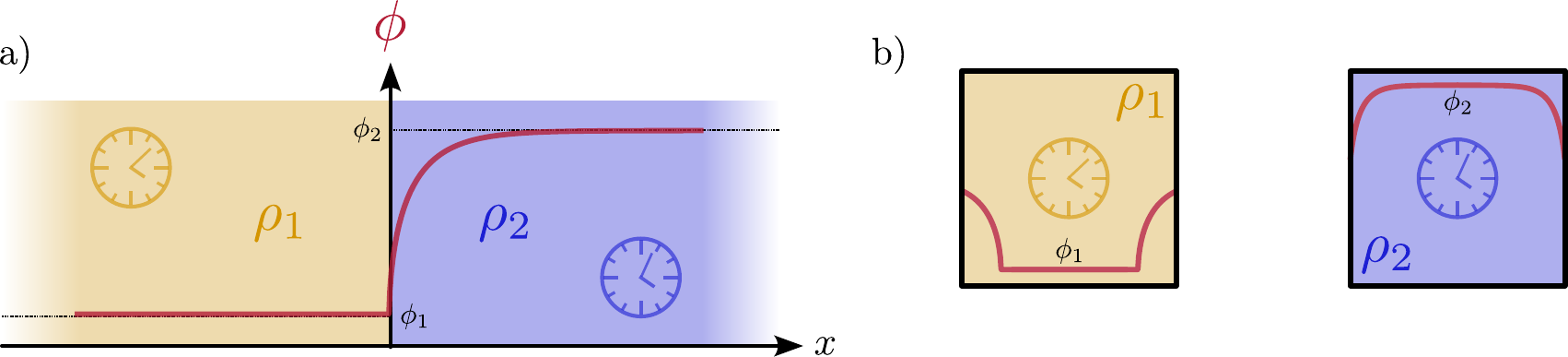}
\caption{A redshift thought experiment with the chameleon model.}
\label{fig:gedankenexperiment}
\end{figure*}

\subsubsection{Setup a)}
\label{subsubsec:setup-a}

Let us start with a basic setup where space is divided into two regions separated by a plane. Region 1 is filled with a fluid of density $\rho_1$ while region 2 is filled with another fluid of density $\rho_2$. In such a configuration, the chameleon field is expected to vary significantly nearby the transition between the two regions, and relax to $\phi_{\mathrm{min}} (\rho_i) \equiv \phi_i$ in the $i^{\mathrm{th}}$ region, $i \in \{1, \, 2\}$. This configuration is depicted in the left-hand panel of Fig.~\ref{fig:gedankenexperiment}, for $\rho_1 \gg \rho_2$. Suppose that we supplement this basic setup with two clocks, one in each of the two regions, placed sufficiently far away from the median plane so that the field does not vary much anymore. Put another way, clock 1 is immersed in a region of space where $\phi = \phi_1$ while clock 2 is immersed in a region of space where $\phi = \phi_2$. We further assume, for now, that the clocks are \textit{perfect} in the sense that they do not perturb the scalar field profile at all.

In this scenario, the two clocks will tick at different frequencies, and the relative frequency shift is given by
\begin{equation}
\begin{split}
    \frac{f_1 - f_2}{f_2} &= \frac{\mathrm{d}\Tilde{\tau}_2}{\mathrm{d}\Tilde{\tau}_1} - 1 = z = \frac{\Omega_2}{\Omega_1} \sqrt{\frac{(g_{00})_2}{(g_{00})_1}} - 1 \\
    &\simeq \Delta_{12} \left( \Phi + \frac{\beta \phi}{M_{\mathrm{Pl}}} \right) \, ,
\end{split}
\label{eqn:clock-ticks1}
\end{equation}
where $\tilde{\tau}_i$ denotes the proper time experienced by the $i^{\mathrm{th}}$ clock. As discussed above, the scalar field has two contributions in this expression: an explicit one, with the $\beta \phi / M_{\mathrm{Pl}}$ term, and a hidden one, with the $\Phi$ term through Eq.~(\ref{eqn:metric-eqn-newtonian-limit-ef}). These two contributions are first discussed analytically. On the one hand, since the scalar field is given by Eq.~(\ref{eqn:chameleon-effective-min}) at the clocks' position (by hypothesis), we have
\begin{equation}
    \Delta_{12} \! \left( \frac{\beta \phi}{M_{\mathrm{Pl}}} \right) = \left[ \left(\frac{\beta}{M_{\mathrm{Pl}}}\right)^{\!n} \! n \Lambda^{n+4} \right]^{\! \frac{1}{n+1}} \!\! \Delta_{12} \! \left(\rho^{-\frac{1}{n+1}}\right)\!.
\label{eqn:scalar-contribution1}
\end{equation}
Eq.~(\ref{eqn:scalar-contribution1}) encapsulates the central idea of this thought experiment. Indeed, it is easy to see that the scalar field contribution can virtually be made as large as one desires in the limit $\rho_2 \to 0$ (for any fixed density $\rho_1$). On the other hand, it is somewhat harder to get an analytical estimate of the scalar field contribution to the potential $\Phi$. Worse, the left panel of Fig.~\ref{fig:gedankenexperiment} depicts two infinite half-spaces of constant density each, so that Eq.~(\ref{eqn:metric-eqn-newtonian-limit-ef}) boils down to
\begin{equation*}
	2 M_{\mathrm{Pl}}^2 \, \Phi''(x) =
	\begin{cases}
		\rho_1 - 2 V \bigl(\phi (x) \bigr) & \text{if } x < 0 \, , \\
		\rho_2 - 2 V \bigl(\phi (x) \bigr) & \text{if } x > 0 \, .
	\end{cases}
\end{equation*}
The lack of obvious physical boundary conditions in this case makes this ODE problem ill-posed.

\subsubsection{Setup b)}
\label{subsubsec:setup-b}

In order to circumvent this thorny issue, we consider a slightly more realistic experimental design where the two clocks are put into separate boxes filled with materials of density $\rho_1$ and $\rho_2$ (in an otherwise vacuum medium), as shown in the right panel of Fig.~\ref{fig:gedankenexperiment}. Provided that the boxes are big enough for the field to reach $\phi_{\min}$ in their interior, the previously exposed qualitative arguments of this thought experiment shall remain valid. Back to the estimation of $\Phi$, the linearity of Eq.~(\ref{eqn:metric-eqn-newtonian-limit-ef}) allows us to decompose this potential as $\Phi = \Phi_N + \Phi_{V}$, where $\Phi_N$ and $\Phi_{V}$ are solutions to
\begin{equation}
	2 M_{\mathrm{Pl}}^2 \, \Delta \Phi_N = \rho \quad \text{and} \quad M_{\mathrm{Pl}}^2 \, \Delta \Phi_{V} = - V(\phi) \, ,
\end{equation}
respectively. By doing so, we can rewrite the total redshift $z$ as $z = z_N + z_{\phi}$ with
\begin{equation}
	z_N = \Delta_{12} \Phi_N \quad \text{and} \quad z_{\phi} = \Delta_{12} \! \left( \frac{\beta \phi}{M_{\mathrm{Pl}}} \right) + \Delta_{12} \Phi_{V} \, .
\label{eqn:redshift-Newtonian-scalar-contributions}
\end{equation}
This decomposition is convenient for physical interpretation because it separates the contribution of the chameleon field from that of the Newtonian potential $\Phi_N$. Yet, we crucially need an estimate of $\Delta_{12} \Phi_{V}$ for comparison against the contribution given by Eq.~(\ref{eqn:scalar-contribution1}). To this end, let us make the simplifying assumptions that \textit{(i)} the boxes are spherical with radius $R_{\mathrm{box}}$ and \textit{(ii)} the boxes are \textit{screened} and exhibit a \textit{thin-shell} of negligible thickness \textemdash \ i.e. the scalar field sits at $\phi_{\min}$ in most of the spherical boxes. Then, using Green's function for the Laplacian at the geometrical center of the boxes yields
\begin{equation}
\begin{array}{r@{\hspace{1\tabcolsep}}l}
\Phi_N  &= \displaystyle - \left( \frac{R_{\mathrm{box}}}{2 M_{\mathrm{Pl}}} \right)^{\! 2} \rho_i\\[0.4cm]
\Phi_{V} &= \displaystyle \frac{1}{2} \! \left( \frac{R_{\mathrm{box}}}{M_{\mathrm{Pl}}} \right)^{\! 2} V \bigl(\phi_{\min} (\rho_i) \bigr)
\end{array}
\ , \quad i \in \{1 , \, 2\} \, .
\end{equation}
With these approximations at hands, we get
\begin{equation}
z_N = - \left( \frac{R_{\mathrm{box}}}{2 M_{\mathrm{Pl}}} \right)^{\! 2} \! \Delta_{12} \rho \, ,
\label{eqn:approx-z-N-boxes}
\end{equation}
\begin{equation}
\begin{split}
z_{\phi} = & \, M_{\mathrm{Pl}}^{- \frac{n}{n+1}} (n \beta^n \Lambda^{n+4})^{\frac{1}{n+1}} \Delta_{12} \! \left(  \rho^{- \frac{1}{n+1}} \right) \\
	&+ \frac{R_{\mathrm{box}}^2}{2} M_{\mathrm{Pl}}^{- \frac{3n+2}{n+1}} (n \beta^n \Lambda^{n+4})^{\frac{1}{n+1}} \Delta_{12} \! \left(  \rho^{\frac{n}{n+1}} \right) .
\end{split}
\label{eqn:approx-z-phi-boxes}
\end{equation}
Recalling that $M_{\mathrm{Pl}} \simeq 2.4 \times 10^{27} \, \mathrm{eV}$ in natural units, we can safely assume that the contribution~(\ref{eqn:scalar-contribution1}) is the dominant term in $z_{\phi}$.\footnote{This has also been numerically verified by computing the two contributions of Eq.~(\ref{eqn:approx-z-phi-boxes}) for the $(\rho_1 , \, \rho_2)$ pairs that we consider in Secs.~\ref{sec:thought-experiment} and \ref{sec:more-realistic-experiments} (see Table~\ref{tab:material-densities}).} Consequently, the $\Delta_{12} \Phi_{V}$ term is not retained in our subsequent analysis. Besides, the Newtonian contribution $z_N$ to the total redshift is not expected to be overwhelmingly larger than $z_{\phi}$, as (\textit{i}) it is weighted by $M_{\mathrm{Pl}}^{-2}$, and (\textit{ii}) the $\Delta_{12} \rho$ term cannot be made as large as $\Delta_{12} \rho^{-1/(n+1)}$ can be.\footnote{The densest materials we can find on Earth have density that do not exceed a few $10^4 \, \mathrm{kg/m^3}$. Conversely, we are able to achieve high vacuum levels in vacuum chambers.}

\subsection{Constraints with finite-size boxes}
\label{subsec:constraints-finite-size-boxes}

A more debatable hypothesis is the one which states that the scalar field indeed reaches $\phi_{\min}$ at the center of the box \textemdash \ as depicted in the right panel of Fig.~\ref{fig:gedankenexperiment}. It is well-known that this situation does not arise when an object is \textit{unscreened}. One relevant quantity to qualitatively assess whether the box is screened (as desired) or not is the Compton wavelength
\begin{equation}
    \lambda_{\phi} (\rho) = \sqrt{\frac{1}{n(n+1) \Lambda^{n+4}} \left( M_{\mathrm{Pl}} \frac{n \Lambda^{n+4}}{\beta \rho} \right)^{\! \frac{n+2}{n+1}}} \, .
\label{eqn:chameleon-compton-wavelength-bis}
\end{equation}
Typically, we would expect the box to have a radius at least a few Compton wavelengths in size, so that our assumption is fulfilled.\footnote{Of course, the Compton wavelength alone is not sufficient to determine whether an object is screened or not. The density of the background medium in which it is embedded must also be taken into account [see e.g. the analytical criterion (\ref{eqn:analytic-screening-criterion})].} Yet, as $\rho_2 \to 0$, $\lambda_{\phi} (\rho_2) \to + \infty$, meaning that the chameleon field would not have enough space to reach $\phi_{\min} (\rho_2)$ within any finite-size box. In Fig.~\ref{fig:thought-experiment-constraints}, we have represented in silver dotted lines the iso Compton wavelength $\lambda_{\phi} (\rho_2) = 1 \, \mathrm{m}$ in the $(\beta^{-1}, \, \Lambda)$-plane, where $\rho_2$ refers to the density of the less dense material of each pair. Above (resp. below) this line, $\lambda_{\phi} (\rho_2) > 1 \, \mathrm{m}$ (resp. $\lambda_{\phi} (\rho_2) < 1 \, \mathrm{m}$). We chose to show the one meter reference as it corresponds to the typical size of objects found in the laboratory. As $\rho_2$ decreases from $\rho_{\mathrm{air}}$ to $\rho_{\textsc{uhv}}$ and to $\rho_{\textsc{ipm}}$, the portion of the parameter space for which the Compton wavelength is smaller than $1 \, \mathrm{m}$ shrinks to the bottom-left corner. In other words, there is a trade-off to be made between (\textit{i}) maximizing $z_{\phi}$ on the one hand, and (\textit{ii}) making sure that the experiment is sensitive to a wide enough area of the parameter space on the other hand. The former condition is an incentive to aim for the best possible level of vacuum for $\rho_2$, while the latter condition requires the two boxes to be sufficiently dense for otherwise the field will not reach $\phi_{\min}$ at their center.

\begin{figure}[t]
\includegraphics[width=0.9\linewidth]{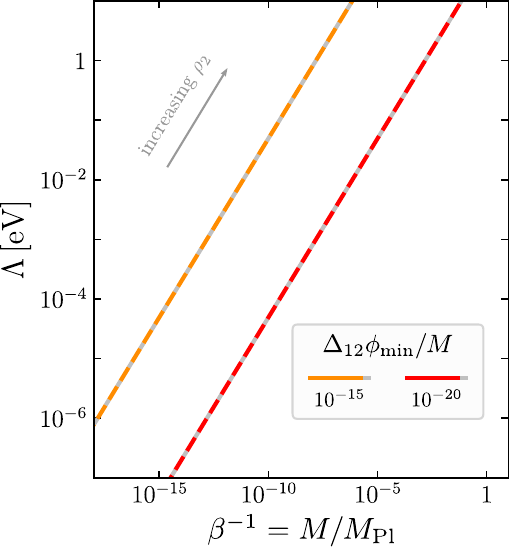}
\caption{Expected constraints on the chameleon from redshift measurements for $\varepsilon_{\mathrm{rel}} \in \{10^{-15} , \, 10^{-20}\}$. This corresponds to revised bounds on parameters $(M , \, \Lambda)$\footref{fn:beta-M} compared to what is presented in Fig.~\ref{fig:thought-experiment-constraints} by accounting for the finite box sizes ($1 \, \mathrm{m}$).}
\label{fig:redshift-constraints-finite-boxes}
\end{figure}

We thus need to revise the forecasts presented in Fig.~\ref{fig:thought-experiment-constraints} by accounting for the fact that the boxes containing the atomic clocks are finite in size. The best solution we found to this constrained optimization problem is to make $\rho_2$ vary continuously, and combine all the resulting constraints together. By doing so, we can derive the best constraints for the relevant $\beta$- and $\Lambda$-ranges. These \textit{weaker} (but more realistic) forecasts for laboratory experiments are obtained by solving the algebraic system
\begin{equation}
\begin{cases}
    \lambda_{\phi} (\rho_2) = R_{\mathrm{box}} = 1 \, \mathrm{m} \\[3pt]
    \beta \Delta_{12} (\phi_{\min}) / M_{\mathrm{Pl}} = \varepsilon_{\mathrm{rel}}
\end{cases} ,
\end{equation}
for $(\beta, \, \Lambda)$, where $\varepsilon_{\mathrm{rel}} \in \{10^{-15} , \, 10^{-20}\}$ denotes the atomic clock relative precision. The resulting bounds, which turn out to be straight lines in $\log$ space, are shown in Fig.~\ref{fig:redshift-constraints-finite-boxes}. These revised bounds exhibit a steeper slope, meaning that high-$M$ (or equivalently, low-$\beta$) regions are more difficult to constrain than what Fig.~\ref{fig:thought-experiment-constraints} suggested.

\section{Scalar field influence on optical energy transitions}
\label{app:energy-transition}

This appendix aims to clarify the influence of the chameleon scalar field on a given energy transition exhibited by strontium-87 isotopes. To this end, we shall use perturbation theory at first order. Several optical lattice clocks take advantage of the ultra-narrow $^1S_0 \to \, ^{3}P_0$ transition \cite{bothwell-2022, zheng-2022, zheng-2023}. In the absence of hyperfine coupling, this specific transition is forbidden due to the selection rules of quantum mechanics. However, $^{87}\text{Sr}$ has a non-zero nuclear spin $I = + 9/2$ which couples to to the electronic angular momentum, making the $^1S_0 \to \, ^{3}P_0$ transition weakly allowed \cite{courtillot-2003}.

\subsection{Atomic Hamiltonian and perturbation theory}

We decompose the total Hamiltonian $\mathcal{H}$ for the $^{87} \text{Sr}$ atom as
\begin{equation}
    \mathcal{H}(\mathbf{r}) = \mathcal{H}_0(\mathbf{r}) + \delta \mathcal{H}_{\text{hc}} (\mathbf{r}) + \delta \mathcal{H}_{\phi} (\mathbf{r}) \, ,
\label{eqn:hamiltonian-decomposition}
\end{equation}
where $\mathcal{H}_0$ is the unperturbed Hamiltonian, $\delta \mathcal{H}_{\text{hc}}$ represents the hyperfine coupling and $\delta \mathcal{H}_{\phi}$ is the scalar field perturbation. Note that $\delta \mathcal{H}_{\text{hc}}$ and $\delta \mathcal{H}_{\phi}$ are assumed to be small perturbations with respect to $\mathcal{H}_0$.

The wavefunctions of the unperturbed energy levels $\psi$ and their associated energy states $E$ are solution to the time-independent Schrödinger equation
\begin{equation}
    \mathcal{H}_0 (\mathbf{r}) \bigl[ \psi(\mathbf{r}) \bigr] = E \psi(\mathbf{r}) \, .
\label{eqn:unperturbed-schrodinger}
\end{equation}
In particular, this provides the wavefunctions of the energy states $\ket{^1S_0}$ and $\ket{^3P_0}$, together with the energy levels $E(^1S_0)$ and $E(^3P_0)$ respectively. We then employ perturbation theory to determine the effects of the perturbing Hamiltonians $\delta \mathcal{H}_{\text{hc}}$ and $\delta \mathcal{H}_{\phi}$.

As mentioned above, the $^1S_0 \to \, ^{3}P_0$ transition is \textit{a priori} not allowed, since $\braket{^1S_0 \, | \, \mathcal{H}_0 \, | \, ^3P_0} = 0$. However, this transition becomes weakly allowed thanks to the hyperfine coupling, which modifies the matrix element as $\braket{^1S_0 \, | \, \delta \mathcal{H}_{\text{hc}} \, | \, ^3P_0} \neq 0$. In addition to enabling this transition, $\delta \mathcal{H}_{\text{hc}}$ also shifts the energy levels ever so slightly.

Let us now take a look at the effects of the scalar field through $\delta \mathcal{H}_{\phi}$. At first order, the energy levels get shifted as
\begin{equation}
\begin{split}
    \Delta E (^1S_0) &= \braket{1^S_0 \, | \, \delta \mathcal{H}_{\phi} \, | \, ^1S_0} \\[5pt]
    \Delta E(^3P_0) &= \braket{^3P_0 \, | \, \delta \mathcal{H}_{\phi} \, | \, ^3P_0}
\end{split}
\label{eqn:shift-scalar}
\end{equation}
Now, the electronic wavefunctions of the states $\ket{^1S_0}$ and $\ket{^3P_0}$ peak several Bohr radii away from the nucleus. Therefore, whatever happens to the scalar field anywhere near the nucleus does not affect the shifts given by Eq.~(\ref{eqn:shift-scalar}). Also, we have shown in Fig.~\ref{fig:nucleus-screening} that the perturbation of the scalar field due to the nucleus does not alter its value at the electron cloud. Therefore, at first order in perturbation theory, the influence of the scalar field on the $^1S_0 \to \, ^{3}P_0$ transition does not depend on whether the nucleus is screened or not. This reasoning remain valid if the atom has zero nuclear spin, as is the case with $^{88}\text{Sr}$.

One may argue that, if the scalar field is uniform across the electronic wavefunction, the two energy levels of interest must be shifted by the same amount so that $\Delta E (^1S_0) - \Delta E(^3P_0) = 0$. At first glance, this uniform shift seems to imply that the transition energy between two levels should remain unchanged. However, the light emitted by such a transition will be subject to the gravitational redshift effect, just as in pure GR except that here $\phi$ contributes to the total gravitational potential [Einstein-frame perspective]. In the case of optical lattice clocks, the frequency of the exciting light must account for this redshift.

\subsection{Higher-order terms}

In the above, we have neglected some physical effects which are expected to be higher-order terms:
\begin{itemize}
    \item[--] We made the assumption that $\phi$ does not vary across the electronic wavefunction of the atom. However, there exists a small dependence on ${r = \| \mathbf{r} \|}$ which should in turn introduce a small shift $\Delta E (^1S_0) - \Delta E(^3P_0) \neq 0$. This is exactly the effect that is pointed out in Refs.~\cite{brax-burrage-2011, brax-davis-2023} which compute bounds on screened scalar field theories from hydrogen-like systems. Here, one should appreciate the fact that such spectroscopic tests are quite different from the redshift measurements proposed in this article.
    \item[--] Second-order perturbations involve cross terms between $\delta \mathcal{H}_{\text{hc}}$ and $\delta \mathcal{H}_{\phi}$. In this case, what happens to the scalar field at the nucleus is expected to become relevant.
\end{itemize}
Finally, it is worth noting that in pure GR, all energy levels get shifted by the same amount, for the metric field hardly varies at all across the atom. In screened scalar-tensor theories however, the fact that the scalar field can vary at scales shorter than that of the atom opens the way to spectroscopic signatures \textemdash \ see e.g. Refs.~\cite{brax-burrage-2011, brax-davis-2023}. In this perspective, it would be interesting to further study the effect of the scalar field on the energy levels of hyperfine substates of electronic states. If the chameleon field influences hyperfine transitions but not optical ones, the redshift measured by atomic clocks would depend on the type of transition they use. In plain language, a pair of clocks based on hyperfine transitions would experience a different redshift than another pair based on optical transitions. Quantifying the magnitude of such an hypothetical effect is beyond the scope of this article.
\clearpage
%

\end{document}